\documentclass[ twoside,journal]{IEEEtran}

    \usepackage{makecell}
    \usepackage{array}
    \usepackage{graphicx,amssymb,amsmath}
    \usepackage{multicol}
    \usepackage[noadjust]{cite}
    \usepackage{setspace}
    \usepackage{subfigure}
    \usepackage{graphicx}
    \usepackage{float}
    \usepackage {url}
    \usepackage{stfloats}
    \usepackage{amsthm,pifont}
    \usepackage{flushend}
    \usepackage{cases,subeqnarray}
    \usepackage{bm,multirow,bigstrut}
    \usepackage{amsmath, amsthm, amssymb}
    \usepackage{textcomp}
    \usepackage{latexsym,bm}
    \usepackage{booktabs}
    \usepackage{mathtools}
    \usepackage{dsfont}
    \usepackage{extarrows}
    \usepackage{epsfig}
    \usepackage{epsfig}
    \usepackage{epstopdf}
    \usepackage[noend]{algpseudocode}
    \usepackage{hyperref}
    \usepackage{algorithmicx,algorithm}

    \usepackage{diagbox} 
    \usepackage{mathrsfs} 
    \usepackage[table]{xcolor} 
    \usepackage{colortbl} 
    \usepackage{multirow, hhline} 
    \usepackage{svg} 

    \theoremstyle{plain}

    \theoremstyle{plain}

    \usepackage{amsmath}

    \usepackage{tablefootnote}
\usepackage{algorithmicx}

    \IEEEoverridecommandlockouts
    \begin{document}
    \title{Generative AI Enabled Robust Data Augmentation for Wireless Sensing in ISAC Networks }
    
    \author{

    Jiacheng Wang, Changyuan Zhao, Hongyang Du, Geng Sun, Jiawen Kang, \\ Shiwen Mao,~\IEEEmembership{Fellow,~IEEE}, Dusit Niyato,~\IEEEmembership{Fellow,~IEEE}, and Dong In Kim,~\IEEEmembership{Life Fellow,~IEEE}
    \vspace{-0.5cm   }
    
    \thanks{Jiacheng Wang, Changyuan Zhao, and Dusit Niyato are with the College of Computing and Data Science, Nanyang Technological University, Singapore (e-mail: jiacheng.wang@ntu.edu.sg, zhao0441@e.ntu.edu.sg, dniyato@ntu.edu.sg).}
    \thanks{Geng Sun is with College of Computer Science and Technology, Jilin University, China 130012, (e-mail: sungeng@jlu.edu.cn).}
    \thanks{Hongyang Du is with the Department of Electrical and Electronic Engineering, University of Hong Kong, Pok Fu Lam, Hong Kong (e-mail: duhy@eee.hku.hk).}
    \thanks{Jiawen Kang is with the School of Automation, Guangdong University of Technology (GDUT), China, (e-mail: kavinkang@gdut.edu.cn).}
     \thanks{Shiwen Mao is with the Department of Electrical and Computer Engineering, Auburn University, Auburn, USA (e-mail: smao@ieee.org)}
    \thanks{Dong In Kim is with the Department of Electrical and Computer Engineering, Sungkyunkwan University, Suwon 16419, South Korea (email:dongin@skku.edu).}
    }
    
\maketitle
    \begin{abstract}
    Integrated sensing and communication (ISAC) uses the same software and hardware resources to achieve both communication and sensing functionalities. Thus, it stands as one of the core technologies of 6G and has garnered significant attention in recent years. In ISAC systems, a variety of machine learning models are trained to analyze and identify signal patterns, thereby ensuring reliable sensing and communications. However, considering factors such as communication rates, costs, and privacy, collecting sufficient training data from various ISAC scenarios for these models is impractical. Hence, this paper introduces a generative AI (GenAI) enabled robust data augmentation scheme. The scheme first employs a conditioned diffusion model trained on a limited amount of collected CSI data to generate new samples, thereby expanding the sample quantity. Building on this, the scheme further utilizes another diffusion model to enhance the sample quality, thereby facilitating the data augmentation in scenarios where the original sensing data is insufficient and unevenly distributed. Moreover, we propose a novel algorithm to estimate the acceleration and jerk of signal propagation path length changes from CSI. We then use the proposed scheme to enhance the estimated parameters and detect the number of targets based on the enhanced data. The evaluation reveals that our scheme improves the detection performance by up to 70\%, demonstrating reliability and robustness, which supports the deployment and practical use of the ISAC network.
    \end{abstract}
    \vspace{-0.2cm}
    \IEEEpeerreviewmaketitle
    \vspace{-0.2cm}
    \section{INTRODUCTION}
    Integrated sensing and communication (ISAC)~\cite{liu2022integrated}, utilizing limited network and hardware resources to support both communication and sensing functions, is one of the fundamental technologies for 6G. To ensure the performance of sensing in ISAC, it is crucial to optimize modulation schemes, waveforms, and other components that allow wireless communication signals to better recording environmental changes~\cite{liu2020radar, liu2020joint}. For instance, the emerging orthogonal time frequency space modulation (OTFS) method modulates data in the Delay-Doppler domain, rather than the traditional time-frequency (TF) domain~\cite{yuan2022orthogonal}. Based on the signal representation in the Delay-Doppler domain, it offers robust Doppler and delay resilience, making it more suitable for the ISAC systems deployed in dynamic environments. 
    
    Besides optimizing the signal itself~\cite{liu2021cramer}, ISAC systems also employ various signal processing algorithms to extract features from signals in the time, frequency, and spatial domains~\cite{wang2023through}. These features are then analyzed and classified through statistical methods, machine learning (ML), and other techniques, thereby realizing effective sensing~\cite{wang2016csi}. For example, the authors in~\cite{wang2024acceleration} estimate the speed and acceleration of signal propagation path length changes, along with the corresponding statistical characteristics, and then use these features to train the support vector machines (SVM) for fall detection. In another study~\cite{chen2017tr}, the authors project channel state information (CSI) into the time reversal resonating strength space and analyze signal features via affinity propagation algorithms. Building on this, they combine automatic label learning with SVM to analyze the extracted signal features for breath detection. These typical CSI-based ISAC systems demonstrate that most sensing tasks rely on signal feature extraction, analysis, and recognition, where various ML-based models plays a vital role~\cite{nirmal2021deep}. These models achieve not only better performance in signal analysis, but also recognition and classification.

    Training ML models for different sensing tasks requires high quality samples~\cite{liu2019wireless}. For instance, in~\cite{zheng2019zero}, to achieve accurate gesture recognition, the authors engaged nearly 20 testers to collect up to 17,000 samples in various settings to train a classifier based on deep learning neural networks. Similarly, in large-scale sensing applications such as activity recognition and localization, training reliable recognition and localization models also requires extensive CSI measurements. For example, in~\cite{he2023robust}, the authors used a dataset containing over 110,000 samples to train a deep learning model based on attention mechanisms, thereby achieving target activity recognition. The authors in~\cite{wang2016csi} collected at least 25,000 samples to train the deep learning model. These studies indicate that a substantial amount of training data is essential for training reliable sensing models for ISAC networks.

    However, collecting training data faces two main challenges. Firstly, to ensure accuracy and reliability of trained ML models, a sufficient amount of high-quality CSI is typically required. This often necessitates the use of specialized equipment or toolkits, such as vector network analyzers or software-defined radio devices~\cite{wang2024aigc}. These devices are costly, complex, and time-consuming to operate, posing challenges to their widespread deployment for collecting training samples across different ISAC environments. Secondly, as most sensing tasks focus on users, such as their locations, activities, and behaviors, collecting CSI typically requires significant user participation. This not only increases labor costs but also raises privacy concerns~\cite{wang2024reflexnoop}, as sensitive private physiological information such as gait, breathing, and heart rate can be extracted from the collected CSI.

    One effective solution to this dilemma is to employ generative AI (GenAI) to augment CSI data for model training~\cite{wang2025aigc}. GenAI models can learn patterns in training samples and then uses the learned knowledge to generate new samples~\cite{wang2024generativeM}. For example, the diffusion model can generate optimal pricing strategies that encourage providers to offer sensing and virtualization services~\cite{wang2024unified}. In ~\cite{hussain2021adaptive}, VAE model is employed to learn probabilistic models of beam dynamics and then the trained VAE is used to design adaptive beam trainers. In ISAC networks, the authors in~\cite{chi2024rf} utilize an improved diffusion model to generate Wi-Fi CSI and frequency modulated continuous wave (FMCW) signals, thereby expanding the dataset to facilitate the training of gesture recognition classifiers. These examples confirm that GenAI models can be used effectively for data augmentation in communication systems and demonstrate that diffusion models possess better generative capabilities~\cite{chi2024rf}, holding potential to address the issue of insufficient training samples in ISAC networks. However, samples generated by diffusion models may not always be optimal, particularly when original training samples are insufficient. In these cases, diffusion models are struggle to capture the distribution characteristics of the original data, leading to the creation of low-quality samples~\cite{dhariwal2021diffusion}.
    
    Essentially, quality of generated samples depends on two key factors, including training samples and computing resources. Firstly, the original samples for training the diffusion model should be of the highest standard in terms of both quality and quantity, allowing the model to learn the distribution and latent patterns~\cite{trabucco2023effective}. Secondly, sufficient computing resources ensure an adequate number of training and inference steps, enabling the model to fully learn the data features and utilize the learned patterns for sample generation through the denoising process~\cite{croitoru2023diffusion}. In various ISAC scenarios, sensing models are trained offline with sufficient computing resources, ensuring that the latent patterns in the training data are thoroughly learned and utilized. However, the original samples for training the diffusion model are often insufficient and unevenly distributed across different labels. In such cases, diffusion models struggle to fully learn the insights of the training samples. Therefore, the samples generated by the trained diffusion model may still be of low quality, making it difficult to support ISAC applications.

    Given the challenges of data augmentation in ISAC systems and capabilities of diffusion models, this paper introduces the Diffusion based Robust data Augmentation scheme (DiRA). Specifically, the proposed DiRA includes the diffusion-based sample quantity expansion and quality enhancement modules. The first module is used to expand the training sample size, while the second module enhances the quality of the generated samples. Moreover, we propose a novel algorithm to estimate the acceleration and jerk of the signal propagation path length change from CSI. These fundamental parameters record the characteristics of signal fluctuations caused by user activities, and can be used in various sensing applications. We employ DiRA to improve the estimated acceleration and jerk. By using the detection of the number of targets as an example, we illustrate that DiRA is more robust and reliable in data augmentation. This is particularly when the original sensing data is insufficient and unevenly distributed. Although we focuses on user detection, DiRA can be adapted to various scenarios, such vehicular networks and UAV-based ISAC systems. The contributions of this paper are summarized as follows.
    \begin{itemize}
    \item We propose a novel algorithm to estimate the acceleration and jerk of the signal propagation path length change from CSI. These parameters record user activities, behaviors, and location changes, providing crucial support for more accurate wireless sensing.

    \item  We propose DiRA, a diffusion based robust sensing data augmentation model, including the quantity expansion module and quality enhancement module. DiRA can generate new sensing samples while enhancing their quality when the original samples for training the diffusion model are insufficient and unevenly distributed. This provides data support for training ML models for various sensing tasks in ISAC systems, when original samples are scarce.

    \item We employ the proposed DiRA to augment the estimated signal parameters and train the ML model with augmented samples for target detection. The results show that using DiRA can increase the average detection accuracy by 0.29,  indicating an improvement of 70\% compared to the case without using it. This confirms that DiRA can effectively enhances data, addressing the problem of insufficient training data in ISAC networks.
\end{itemize}
\vspace{-0.3cm}
   \section{Related Work}
   This section discusses the existing works about CSI-based wireless sensing and GenAI-based data augmentation.
   \vspace{-0.4cm}
   \subsection{CSI-based Wireless Sensing}
   CSI-based ISAC systems have developed rapidly in recent years, with researchers proposing various algorithms to extract signal features and integrate them with diverse analysis and recognition methods for different sensing tasks. Common features for target localization include signal angle of arrival (AoA), time of flight (ToF), angle of departure (AoD), and signal attenuation. For instance, the authors in~\cite{yang2021decimeter} estimate signal AoA and ToF from CSI and then build spatial constrains to realize passive human target localization. Similarly, the authors in~\cite{zheng2019zero} introduce a unified model that considers AoA, ToF, and Doppler frequency shift, and devises an efficient algorithm to jointly estimate them. After refining these parameters of the target induced reflected signals, the authors then incorporate transceiver locations to achieve passive target localization. Besides localization, these signal parameters can also be used for target detection and tracking~\cite{li2024wifi}.

    Moreover, some other studies extract more signal parameters such as Doppler shifts and velocity profiles from CSI, which can support sensing tasks such as behavior recognition, fall detection, and fine-grained gesture recognition~\cite{wang2017phasebeat}. For example, the authors in~\cite{wang2020csi} demonstrate that in indoor multipath environments with small-scale fading, the CSI phase difference is a periodic signal with the same frequency as breathing when the signal is reflected from the chest of a person. Hence, using the phase difference between receiver antennas, they realize accurately respiratory and heart rate detection. Besides, authors in~\cite{niu2021understanding} develop a model to quantify the relationship between signal frequency and the target location, movement direction, and velocity of human motion. Based on this, they show that speed and direction are location-dependent, while movement fragments and relative motion direction changes are location-independent. Therefore, leveraging these findings, they achieve location-independent gesture recognition. Unlike existing works, this paper proposes an algorithm to estimate the acceleration and jerk of signal propagation path length changes. These parameters provide critical support for sensing tasks, such as detecting the number of targets and evaluating their movement characteristics.

        \begin{figure*}[htp]
    \centering
    \includegraphics[width=1\textwidth]{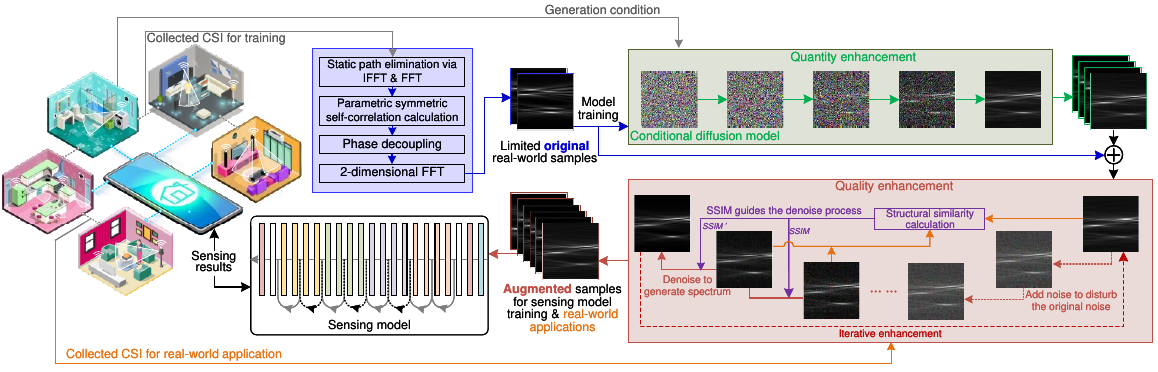}
    \caption{The overview of the proposed scheme. It first extracts spectra from the CSI collected from real-world environments, and then augments the limited signal spectra using DiRA. After that, the augmented signal spectra are used to train the sensing model, thereby achieving the detection of the number of targets. Here, DiRA includes the sample quantity expansion and quality enhancement module. The first one contains a conditional diffusion model, which is trained on a limited number of signal spectra collected in real-world ISAC scenarios. During the offline phase, it can generate signal spectra based on input conditions, thereby achieving sample quantity expansion. The second one is a diffusion model trained on noise-free signal spectra, used to further enhance the quality of the signal spectra. Note that during the online phase, the input sensing data must pass through the quality enhancement module before going to the sensing model to ensure consistency.}
    \label{SYS}
    \vspace{-0.5cm}
    \end{figure*}

   \subsection{GenAI-based Signal Augmentation}
   Research on GenAI-based signal enhancement primarily focuses on VAE, GAN, and diffusion models~\cite{zhou2021indoor}. For instance, authors in~\cite{zhang2024vawss} utilize the VAE to learn the latent representations of CSI amplitudes and then generate CSI amplitude for different subcarriers, which can simulate various communication scenarios. In~\cite{kompella2024augmenting}, the authors propose a vector-quantized variational autoencoder (VQ-VAE), capable of synthesizing high-fidelity radio frequency (RF) signals by capturing the inherent complex variations in RF communication signals, thereby enhancing the diversity and fidelity of the training dataset. Such enhancement is particularly useful under low SNR conditions, as it can improve the generalization and robustness of classifiers when collected data is limited.

    Additionally, GANs are also used for signal generation. For instance, authors in~\cite{zou2020adversarial} use GAN to generate received signal strength for areas without signal coverage, enabling the simulation of signal distributions in complex indoor environments, which reduces the time and effort required to build radio maps. In~\cite{zhao2021gsmac}, the authors propose a GAN-based active signal map reconstruction method (GSMAC), which can update the signal map with only a few observations while utilizing incomplete historical signals to update the signal map online. Besides VAE and GAN, in~\cite{chi2024rf}, the authors proposed a time-frequency diffusion model and introduced a hierarchical diffusion transformer, which can produce high-quality time-series RF data through optimizations of the network architecture, functional modules, and complex-valued operators.

    The aforementioned works demonstrate that GenAI can effectively generate various types of signal data, thereby supporting communication and sensing functions. However, these studies overlook that GenAI models trained with insufficient original samples might generate lower-quality data. Therefore, this paper introduces DiRA, which increases data quantity while improving the quality of the generated data, offering more robust support for ISAC networks.
    
   \section{System Model}
   This section introduces the proposed DiRA. Firstly, we introduce how to estimate the acceleration and its change rate of the propagation path length change from raw CSI. Subsequently, using the estimation results as inputs, we present how the proposed DiRA, including the data sample quantity and quality enhancement modules, augments the samples.

   \subsection{Signal Parameter Estimation}
   We consider a wireless device communicates with another device via orthogonal frequency division multiplexing (OFDM) signals. During the communication, the signal receiver uses the surveillance and reference channels to capture the wireless signals and extracts the CSI to sense moving targets in the environment. Following the modeling approach presented in existing works~\cite{wang2024generative,wang2024acceleration}, the conjugate multiplication is conducted between these two channels to eliminate the synchronization errors. Hence, at time zero, the CSI corresponding to the $k$-th subcarrier obtained by the receiver can be denoted as 
    \begin{align}\label{eq1}
    {H_k}\left( 0 \right) = \sum\limits_{l = 1}^L {{A_{k,l}}} \exp \left( {j2\pi {\theta _{k,l}}} \right) + {H_{s\_k}} + {n_k}\left( 0 \right),
    \end{align}    
    where ${A_{k,l}}$ is the amplitude, $L$ is the total number of propagation paths caused by dynamic targets, ${\theta _l}$ is the initial phase of the $l$-th propagation path, ${H_{s\_k}}$ is the sum of the CSI corresponding to the static propagation paths, and ${n_k}\left( 0 \right)$ is the noise. We consider that the $l$-th propagation path is introduced by the dynamic target, and ${v_l}$, ${a_l}$, and ${r_l}$ denote the corresponding velocity, acceleration, and jerk of the propagation path length change, respectively. Then, the path length change over a CSI sampling interval of the $l$-th propagation path is
    \begin{align}\label{eq2}
    {d_l}(\Delta t) = {v_l}\Delta t + \frac{{{a_l}}}{2}\Delta {t^2} + \frac{{{r_l}}}{6}\Delta {t^3}.
    \end{align}    
    The phase accumulation caused by this length change can be represented as ${{{d_l}(\Delta t){f_k}} \mathord{\left/{\vphantom {{{d_l}(\Delta t){f_k}} c}} \right.\kern-\nulldelimiterspace} c}$, where ${f_k}$ is the frequency corresponding to the $k$-th subcarrier and $c$ is the speed of electromagnetic wave propagation. Taking the CSI at time zero in (\ref{eq1}) as the reference, at time $\Delta t$, we have 
    \begin{align}\label{eq3}
{H_k}\left( {\Delta t} \right) &= \sum\limits_{l = 1}^L {{A_{k,l}}\exp \left\{ {j2\pi \left[ {{\theta _{k,l}} + \frac{{{f_k}{d_l}(\Delta t)}}{c}} \right]} \right\}} \\ \notag
 &+ {H_{s\_k}} + n\left( {\Delta t} \right)\\ \notag
 &= \sum\limits_{l = 1}^L {{A_{k,l}}\exp \left[ {j2\pi \left( {{\theta _{k,l}} + \frac{{{v_l}}}{c}{f_k}\Delta t} \right.} \right.} \\ \notag
&\left. {\left. { + \frac{{{a_l}}}{{2c}}{f_k}\Delta {t^2} + \frac{{{r_l}}}{{6c}}{f_k}\Delta {t^3}} \right)} \right] + {H_{s\_k}} + {n_k}\left( {\Delta t} \right).
    \end{align}

   From the analysis above, we can see that the movement of the dynamic target introduces changes in the propagation path length, which is reflected as the phase accumulation of the CSI. Meanwhile, ${H_{s\_k}}$ includes static propagation paths, such as the direct path from the transmitter to the receiver. Hence, it has stronger strength and the corresponding phase accumulation is zero, as the static path lengths do not change over time. To estimate acceleration and jerk of the propagation paths introduced by dynamic target more accurately, it is necessary to eliminate the impact of ${H_{s\_k}}$.

    Therefore, we use $Q$ CSI samples to construct the vector $\left[ {{H_k}\left( 0 \right), \cdots ,{H_k}\left( {\left( {Q - 1} \right)\Delta t} \right)} \right]$ and transform it into the frequency domain via fast Fourier transform (FFT). As the phase of the static paths in ${H_{s\_k}}$ does not accumulate over time, we modify the value corresponding to the zero-frequency component to zero so as to eliminate impact of ${H_{s\_k}}$. On this basis, the modified vector is transformed back into the time domain through inverse FFT. Let ${{\bf{H}}_k}$ represent the vector transformed back to the time domain, based on which we further estimate the dynamic path parameters ${a_l}$ and ${r_l}$. Concretely, we first define the parametric symmetric self-correlation function as follows:
    \begin{align}\label{eq4}
    {{\bf{R}}_{{H_k}}}\left( {\Delta t,\tau } \right) &= \left[ {{{\bf{H}}_k}\left( {\Delta t + B} \right) \times {{\bf{H}}_k}\left( {\Delta t - B} \right)} \right]\\ \notag
    &\times {\left[ {{{\bf{H}}_k}\left( {\Delta t + B'} \right) \times {{\bf{H}}_k}\left( {\Delta t - B'} \right)} \right]^ * },
    \end{align}
where
    \begin{align}\label{eqxx}
    \left\{ \begin{array}{l}
B = \tau  + \varphi  + {{{\tau _d}} \mathord{\left/
 {\vphantom {{{\tau _d}} 2}} \right.
 \kern-\nulldelimiterspace} 2}\\
B' = \tau  - \varphi  + {{{\tau _d}} \mathord{\left/
 {\vphantom {{{\tau _d}} 2}} \right.
 \kern-\nulldelimiterspace} 2}.
\end{array} \right.
    \end{align}
$\tau $ is lag time variable, ${\left(  \cdot  \right)^*}$ is the complex conjugation operator, ${\tau _d}$ represents a constant delay, and $\varphi$ can be defined as $0.089Q$ according to~\cite{djurovic2012hybrid}. Then, substituting~(\ref{eq3}) into (\ref{eq4}), we have 
    \begin{align}\label{eq5}
{{\bf{R}}_{{H_k}}}\left( {\Delta t,\tau } \right) &= \sum\limits_{l = 1}^L {A_{k,l}^4\exp \left\{ {j2\pi } \right.} \left[ {2\frac{{{a_l}}}{c}{f_k}\varphi \left( {2\tau  + {\tau _d}} \right)} \right.\\ \notag
&\left. {\left. { + 2\frac{{{r_l}}}{c}{f_k}\varphi \Delta t\left( {2\tau  + {\tau _d}} \right)} \right\}} \right]\\ \notag
 &+ {{\bf{R}}_{{H_k},cross}}\left( {\Delta t,\tau } \right),
    \end{align}
where the first term represent the self-term of each dynamic signals introduced by the dynamic target, and ${{\bf{R}}_{{H_k},cross}}\left( {\Delta t,\tau } \right)$ is the cross-term. Here, we exclude the noise term to simplify the derivation.

From (\ref{eq5}), we can observe that $\Delta t$ and $\tau $ are coupled in the exponential phase term. Therefore, performing the FFT along the $\Delta t$ axis results in the energy of the self-term peaking along the ${f^{[\Delta t]}} = {{2{r_l}{f_k}} \mathord{\left/
 {\vphantom {{2{r_l}{f_k}} c}} \right.
 \kern-\nulldelimiterspace} c}\left( {2\tau  + {\tau _d}} \right)$, where ${f^{[\Delta t]}}$ represents the frequency domain with respect to $\Delta t$. This implies that, after FFT along the $\Delta t$, another FFT along the $\tau $ axis cannot accumulate the signal energy of the self-term. Hence, inspiring by the previous work~\cite{lv2011lv}, we employ the keystone transformation to decouple these elements, which is defined as follows
    \begin{align}\label{eq6}
    \Delta t = \frac{\eta}{{s\left( {2\tau  + {\tau _d}} \right)}},
    \end{align}
where $\eta$ is the scaled time, $s$ is the scaling factor, and $s{\tau _d} = 1$ according to \cite{lv2009keystone}. Conducting the defined transformation on (\ref{eq5}), we have 
    \begin{align}\label{eq7}
{{\bf{R'}}_{{H_k}}}\left( {\eta,\tau } \right) &= \sum\limits_{l = 1}^L {A_{k,l}^4} \exp \left\{ {j2\pi \left[ {2\frac{{{a_l}}}{c}{f_k}\varphi \left( {2\tau  + {\tau _d}} \right)} \right.} \right.\\ \notag
&\left. {\left. { + 2\frac{{{r_l}}}{c}{f_k}\varphi \frac{\eta}{s}} \right]} \right\} + {{R'}_{{H_k},cross}}\left( {\eta,\tau } \right).
    \end{align}
Comparing (\ref{eq5}) and (\ref{eq7}), we can observe the coupling between $\Delta t$ and $\tau $ is resolved. On this basis, the FFT is conducted along $\eta$, obtaining 
    \begin{align}\label{eq8}
{{\bf{P}}_t}\left( {{f^{\left[ \eta \right]}},\tau } \right) &= {\mathscr{F}_\eta}\left[ {{{\bf{R'}}_{{H_k}}}\left( {\eta,\tau } \right)} \right]\\ \notag
 &= \sum\limits_{l = 1}^L {A_{k,l}^4} \exp \left\{ {j2\pi \left[ {2\frac{{{a_l}}}{c}{f_k}\varphi \left( {2\tau  + {\tau _d}} \right)} \right.} \right.\\ \notag
 &+ \left. {\left. {\delta \left( {{f^{\left[ \eta \right]}} - \frac{{2{r_l}{f_k}\varphi }}{{sc}}} \right)} \right]} \right\} + {P_{\eta,cross}}\left( {{f^{\left[ \eta \right]}},\tau } \right),
    \end{align}
where ${\mathscr{F}_\eta}\left(  \cdot  \right)$ denotes the FFT operation along $\eta$, ${f^{\left[ \eta \right]}}$ is the frequency space corresponding to $\eta$, $\delta \left(  \cdot  \right)$ is the Dirac delta function, and ${{\bf{P}}_{\eta,cross}}\left( {{f^{\left[ \eta \right]}},\tau } \right)$ is the FFT result of the cross-term along the $\eta$ axis. 

From (\ref{eq8}), after decoupling and performing FFT along the $\eta$ axis, the signal energy of each component accumulates into a line ${f^{\left[ \eta \right]}} = {{2{r_l}{f_k}\varphi } \mathord{\left/
 {\vphantom {{2{r_l}{f_k}\varphi } sc}} \right.
 \kern-\nulldelimiterspace}sc}$. After that, anther FFT is conducted on (\ref{eq8}) along the $\tau$ axis, thereby further accumulating the signal energy and obtaining the acceleration-jerk (AJ) spectrum 
     \begin{align}\label{eq9}
{\bf{P}}\left( {{f^{\left[ \eta \right]}},{f^{\left[ \tau  \right]}}} \right) &= {\mathscr{F}_\tau }\left[ {{{\bf{P}}_t}\left( {{f^{\left[ \eta \right]}},\tau } \right)} \right]\\ \notag
 &= \sum\limits_{l = 1}^L {A_{k,l}^4} \exp \left( {j4\pi \frac{{{a_l}}}{c}{f_k}\varphi {\tau _d}} \right)\\  \notag
 &\times \delta \left( {{f^{\left[ \tau  \right]}} - 4\frac{{{a_l}}}{c}{f_k}\varphi } \right)\delta \left( {{f^{\left[ \eta \right]}} - \frac{{2{r_l}{f_k}\varphi }}{{sc}}} \right)\\ \notag
 &+ {{\bf{P}}_{cross}}\left( {{f^{\left[ \eta \right]}},{f^{\left[ \tau  \right]}}} \right),
    \end{align}
where ${f^{\left[ \tau  \right]}}$ is the frequency space corresponding to $\tau$, ${{\bf{P}}_{cross}}\left( {{f^{\left[ \eta \right]}},{f^{\left[ \tau  \right]}}} \right)$ is the FFT result of cross-term. As shown in (\ref{eq9}), it can be seen that after performing FFT across two dimensions, each self-term can be transformed into a peak in the AJ spectrum. Meanwhile, due to the nonlinearity of the parameter-symmetric autocorrelation function, cross-terms also appear in the AJ spectrum. Concretely, according to \cite{djurovic2012hybrid,wang2024acceleration}, the parametric symmetric self-correlation function can expressed as 
     \begin{align}\label{eq10}
{{\bf{R}}_{{H_k}}}\left( {\Delta t,\tau } \right) = {\bf{R}}\left( {\Delta t + \tau  + \frac{{{\tau _d}}}{2}} \right){{\bf{R}}^*}\left( {\Delta t - \tau  - \frac{{{\tau _d}}}{2}} \right),
    \end{align}
where 
     \begin{align}\label{eq11}
{\bf{R}}\left( {\Delta t} \right) = {{\bf{H}}_k}\left( {\Delta t + \varsigma } \right){\bf{H}}_k^*\left( {\Delta t - \varsigma } \right).
    \end{align}
    
From (\ref{eq10}) and (\ref{eq11}), we can observe that in the presence of multiple dynamic paths, if the cross-terms of ${\bf{R}}\left( {\Delta t} \right)$ resemble the form of linear frequency modulation signals, then according to~\cite{lv2011lv}, the cross-terms in ${{\bf{P}}_{cross}}\left( {{f^{\left[ \eta \right]}},{f^{\left[ \tau  \right]}}} \right)$ can accumulate as self-terms. In such a case, the cross-term will also appear in ${\bf{P}}\left( {{f^{\left[ \eta \right]}},{f^{\left[ \tau  \right]}}} \right)$ in the form of Dirac delta function, thereby affecting the detection of the peak points corresponding to the real self-terms. Fortunately, the above situation is almost impossible to happen in the proposed method, which is discussed below.

We conduct a specific analysis via the case with $ L = 2 $. For other situations, we can analyze in a similar way. Specifically, when $ L = 2 $, we have 
     \begin{align}\label{eq12}
H_k^{\left( 2 \right)}\left( {\Delta t} \right) &= {A_{k,1}}\exp \left[ {j2\pi \left( {{\theta _{k,1}} + \frac{{{v_1}}}{c}{f_k}\Delta t} \right.} \right.\\ \notag
 &+ \left. {\left. {\frac{{{a_1}}}{{2c}}{f_k}\Delta {t^2} + \frac{{{r_1}}}{{6c}}{f_k}\Delta {t^3}} \right)} \right]\\ \notag
& + {A_{k,2}}\exp \left[ {j2\pi \left( {{\theta _{k,2}} + \frac{{{v_2}}}{c}{f_k}\Delta t} \right.} \right.\\ \notag
&\left. {\left. { + \frac{{{a_2}}}{{2c}}{f_k}\Delta {t^2} + \frac{{{r_2}}}{{6c}}{f_k}\Delta {t^3}_k} \right)} \right].
    \end{align}
Building on this, by substituting (\ref{eq12}) into (\ref{eq11}), we can obtain
     \begin{align}\label{eq13}
    {R_{cross}}\left( {\Delta t} \right) = {A_{k,12}} \cdot \Phi  + {A'_{k,12}} \cdot \Phi',
    \end{align}
where
     \begin{align}\label{eq14}
    {A_{k,12}} &= {A_{k,1}}{A_{k,2}}\exp \left[ {\frac{{j2\pi }}{{{\lambda _k}}}\left( {{v_1}\varphi  + {v_2}\varphi  + \frac{1}{2}{a_1}{\varphi ^2}} \right)} \right]\\ \notag
    &\times \exp \left[ {\frac{{j\pi }}{{{\lambda _k}}}\left( { - {a_2}{\varphi ^2} + \frac{1}{3}{r_1}{\varphi ^3} + \frac{1}{3}{r_2}{\varphi ^3}} \right)} \right],
    \end{align}
    
    \begin{align}\label{eq15}
    \Phi  &= \exp \left[ {\frac{{j2\pi }}{{{\lambda _k}}}\left( {{v_1} - {v_2} + {a_1}\varphi  + {a_2}\varphi  + \frac{{{r_1}}}{2}{\varphi ^2} - \frac{{{r_2}}}{2}{\varphi ^2}} \right)\Delta t} \right]\\ \notag
    &\times \exp \left[ {\frac{{j\pi }}{{{\lambda _k}}}\left( {{a_1} - {a_2} + {r_1}\varphi  + {r_2}\varphi } \right)\Delta {t^2}} \right]\\ \notag
    &\times \exp \left[ {\frac{{j\pi }}{{3{\lambda _k}}}\left( {{r_1} - {r_2}} \right)\Delta {t^3}} \right],
    \end{align}
    
    \begin{align}\label{eq16}
    {{A'}_{k,12}} &= {A_{k,1}}{A_{k,2}}\exp \left[ {\frac{{j2\pi }}{{{\lambda _k}}}\left( {{v_1}\varphi  + {v_2}\varphi  + \frac{1}{2}{a_2}{\varphi ^2}} \right)} \right]\\ \notag
    &\times \exp \left[ {\frac{{j\pi }}{{{\lambda _k}}}\left( { - {a_1}{\varphi ^2} + \frac{1}{3}{r_1}{\varphi ^3} + \frac{1}{3}{r_2}{\varphi ^3}} \right)} \right],
    \end{align}

    \begin{align}\label{eq17}
    \Phi'  &= \exp \left[ {\frac{{j2\pi }}{{{\lambda _k}}}\left( {{v_2} - {v_1} + {a_1}\varphi  + {a_2}\varphi  + \frac{{{r_2}}}{2}{\varphi ^2} - \frac{{{r_1}}}{2}{\varphi ^2}} \right)\Delta t} \right]\\ \notag
    &\times \exp \left[ {\frac{{j\pi }}{{{\lambda _k}}}\left( {{a_2} - {a_1} + {r_1}\varphi  + {r_2}\varphi } \right)\Delta {t^2}} \right]\\ \notag
    &\times \exp \left[ {\frac{{j\pi }}{{3{\lambda _k}}}\left( {{r_2} - {r_1}} \right)\Delta {t^3}} \right],
    \end{align}
and ${\lambda _k} = {c \mathord{\left/  {\vphantom {c {{f_k}}}} \right.  \kern-\nulldelimiterspace} {{f_k}}}$ is the signal wavelength. From the derivation process described above, it is clear that the cross-terms of ${\bf{R}}\left( {\Delta t} \right)$ have the same form as linear frequency modulation signals only when ${r_1} = {r_2}$, thereby affecting the detection of self-terms. However, in practice, the user's location is unknown and their moving direction and speed are random, encountering ${r_1} = {r_2}$ rarely happen. Therefore, it can be concluded that ${\bf{R}}\left( {\Delta t} \right)$ does not take the form of linear frequency modulation signals, and hence will not affect the detection of peaks corresponding to the self-terms.

Building on this, for each peak in the AJ spectrum corresponding to paths introduced by dynamic targets, the coordinates are given as 
     \begin{align}
\left\{ \begin{array}{l}
{f^{\left[ \eta \right]}} = 4\frac{{{a_l}{f_k}}}{c}\varphi \\
{f^{\left[ \tau  \right]}} = 2\frac{{{r_l}{f_k}}}{{sc}}\varphi 
\end{array} \right..
    \end{align}
Based on these coordinates, we can calculate the acceleration and jerk of the signal propagation path length changes as
 
     \begin{align}
\left\{ \begin{array}{l}
{a_l} = \frac{{{f^{\left[ \eta \right]}}c}}{{4{f_k}\varphi }}\\
{r_l} = \frac{{{f^{\left[ \tau  \right]}}cs}}{{2{f_k}\varphi }}
\end{array} \right..
    \end{align}
 
 These parameters provide crucial support for applications such as fall detection. Moreover, considering that different targets have distinct accelerations and jerks, the number of peaks in the AJ spectrum also reflects the number of dynamic targets, hence supporting the detection of the number of targets.

\subsection{Spectrum Augmentation}
While the obtained AJ spectra can be used for various sensing tasks, in real-world applications, it may not be possible to obtain sufficient AJ spectrum samples for AI training models, such as ML-based classifiers. Therefore, we propose to use GenAI to enhance the AJ spectrum. Based on the collected AJ spectra, we first train the conditional diffusion model to expand the data quantity. On this basis, the AJ spectra generated by the trained conditional diffusion model may be noisy when the original training AJ spectrum samples are insufficient. Hence, we further purify the generated spectra, thereby improving the quality of the augmented AJ samples.
\subsubsection{Spectrum Quantity Expansion}
We employ the conditional diffusion model for the AJ spectrum quantity expansion. The diffusion model consists of two Markov chains, i.e., the diffusion process and the reverse process. The diffusion process gradually adds noise to the AJ spectrum until it becomes Gaussian noise, while the reverse process iteratively denoises the sampled Gaussian noise to generate new AJ spectrum. Let ${{\bf{P}}_0}$ represent ${\bf{P}}\left( {{f^{\left[ \eta \right]}},{f^{\left[ \tau  \right]}}} \right)$ in (\ref{eq9}) for training, then the diffusion process from ${{\bf{P}}_0}$ to ${{\bf{P}}_T}$ can be presented as
    \begin{align}\label{eq18}
    q\left( {\left. {{{\bf{P}}_1}, \ldots ,{{\bf{P}}_T}} \right|{{\bf{P}}_0}} \right) = \prod\limits_{t = 1}^T {{\cal N}\left( {{{\bf{P}}_t};\sqrt {1 - {\beta _t}} {{\bf{P}}_{t - 1}},{\beta _t}{\bf{I}}} \right)},
    \end{align}
where ${\beta _t}$ is the pre-defined positive constants. Let ${\alpha _t} = 1 - {\beta _t}$ and ${\bar \alpha _t} = \prod\nolimits_{i = 1}^t {{\alpha _i}} $, then 
    \begin{align}\label{eq19}
    q\left( {\left. {{{\bf{P}}_T}} \right|{{\bf{P}}_0}} \right) = {\cal N} \left( {{{\bf{P}}_t};\sqrt {{{\bar \alpha }_t}} {{\bf{P}}_0},\left( {1 - {{\bar \alpha }_t}} \right){\bf{I}}} \right).
    \end{align}

Therefore, when $T$ is sufficiently large, ${\bar \alpha _t}$ approaches zero and $q\left( {\left. {{{\bf{P}}_T}} \right|{{\bf{P}}_0}} \right)$ becomes close to the standard normal distribution. During this process, ${{\bf{P}}_T}$ can be obtained through
    \begin{align}\label{eq20}
{{\bf{P}}_t} = \sqrt {{{\bar \alpha }_t}} {{\bf{P}}_0} + \sqrt {\left( {1 - {{\bar \alpha }_t}} \right)} {{\bm{\varepsilon }}_0},
    \end{align}
where ${{\bm{\varepsilon }}_0}$ is the standard Gaussian noise. The reverse process is another Markov process that predicts and eliminates the noise added during the diffusion process, thereby enabling AJ spectrum generation. However, in practice, $q\left( {\left. {{{\bf{P}}_{t-1}}} \right|{{\bf{P}}_{t}}} \right)$ is difficult to compute. Hence, we use a parameterized model ${p_\theta }\left( {\left. {{{\bf{P}}_{t - 1}}} \right|{{\bf{P}}_t}} \right)$ to estimate it. Consequently, the reverse process is defined as
    \begin{align}\label{eq21}
{p_\theta }\left( {\left. {{{\bf{P}}_0}, \ldots ,{{\bf{P}}_{T - 1}}} \right|{{\bf{P}}_T}} \right) &= \prod\limits_{t = 1}^T {{p_\theta }\left( {\left. {{{\bf{P}}_{t - 1}}} \right|{{\bf{P}}_t}} \right)} \\ \notag
 &= \prod\limits_{t = 1}^T {{\cal N}\left( {{{\bf{P}}_{t - 1}};{{\bf{\mu }}_\theta }\left( {{{\bf{P}}_t},t} \right),{\Sigma _\theta }\left( {{{\bf{P}}_t},t} \right)} \right)}, 
    \end{align}
where ${{\bf{\mu }}_\theta }\left( {{{\bf{P}}_t},t} \right)$ and ${\Sigma _\theta }\left( {{{\bf{P}}_t},t} \right)$ are the mean and variance of the parameterized model ${p_\theta }$, respectively. Based on this model, we sample ${{\bf{P}}_T} \sim {\cal N}\left( {{\bf{0,I}}} \right)$ and then calculate ${{\bf{P}}_{t - 1}} \sim {p_\theta }\left( {\left. {{{\bf{P}}_{t - 1}}} \right|{{\bf{P}}_t}} \right)$ for $t = T, \ldots ,1$ to generate new AJ spectrum. 

Given that the AJ spectra obtained under different practical ISAC conditions vary, we use the conditional diffusion model for AJ spectrum generation. Therefore, during the training process, paired samples $\left( {{{\bf{P}}_0},{{\bf{c}}_0}} \right)$, where ${\bf{c}}_0$ is the generation condition\footnote{The AJ spectrum can be utilized for various ISAC applications, such as detecting the number of targets and fall detection. Therefore, the conditions can be set based on the specific application.}, are used to train the conditional diffusion model. After that, the reverse process is modified into a conditional reverse process
    \begin{align}\label{eq22}
{p_\theta }\left( {{{\bf{P}}_{0:T}}\left| {\bf{c}} \right.} \right) = {p_\theta }\left( {{{\bf{P}}_T}} \right)\prod\limits_{t = 1}^T {{p_\theta }\left( {{{\bf{P}}_{t - 1}}|{{\bf{P}}_t},{\bf{c}}} \right)}, 
    \end{align}
where ${p_\theta }\left( {{{\bf{P}}_{t - 1}}|{{\bf{P}}_t},{\bf{c}}} \right) = {\cal N}\left( {{{\bf{P}}_{t - 1}};{{\bf{\mu }}_\theta }\left( {{{\bf{P}}_t},t,{\bf{c}}} \right),{{\bf{\Sigma }}_\theta }\left( {{{\bf{P}}_t},t,{\bf{c}}} \right)} \right)$ and ${\bf{c}}$ is the condition. Here, the loss function is defined as 
    \begin{align}\label{eq23}
{\cal L}\left( \theta  \right) \propto \left\| {\varepsilon \left( {{x_t},t} \right) - {\varepsilon _\theta }\left( {{x_t},t,{\bf{c}}} \right)} \right\|.
    \end{align}
    
For datasets with a limited number of AJ spectra, the trained conditional diffusion model is used to generate more spectra, so as to increase the overall quantity. However, due to the limited original training samples, the conditional diffusion models may struggle to capture the characteristics of the AJ spectra. Hence, the generated spectra are noisy, making them difficult to support ISAC applications such as target activity recognition, which rely on the trained ML-based classifiers. 

\subsubsection{Spectrum Quality Enhancement}
To enhance the quality of the generated AJ spectra, we further propose sample enhancement based on the diffusion model. The core idea here is that the diffusion process can inject Gaussian noise into the generated noisy AJ spectra to overwhelm the inherent noise. Then, the reverse process can eliminate both the inherent noise in the generated spectra and the noise introduced during the diffusion process, thereby achieving spectrum enhancement. The model used here is another diffusion model trained with AJ spectra without noise\footnote{The data can be obtained through simulation.}. Therefore, during the reverse process, it tends to denoise the AJ spectra towards a noise-free domain. Specifically, by assuming that the noisy AJ spectrum generated during the data augmentation process is ${{\bf{P'}}_c}$, then by adding noise via ${T^*}$ steps, we can obtain
    \begin{align}\label{eq24}
{{\bf{P'}}_{{T^*}}} = \sqrt {{{\bar \alpha '}_{{T^*}}}} \left( {{{{\bf{P'}}}_c}} \right) + \sqrt {1 - {{\bar \alpha '}_{{T^*}}}} {{\bm{\varepsilon }}_0},
    \end{align}
where ${\bar \alpha '_{{T^*}}}$ is the parameter of added noise. 

Let ${{\bf{P'}}_c} = {{\bf{P}}_{zr}} + {\bm{\kappa }} $, where ${\bm{\kappa }} $ represents noise, and ${{\bf{P}}_{zr}}$ is the corresponding spectrum without noise, then an appropriate threshold ${T^*}$ need to be selected to allow the injected Gaussian noise to perturb ${\bm{\kappa }} $ while ensuring that the characteristics of ${{\bf{P}}_{zr}}$ are preserved. On this basis, in the reverse process, ${{\bf{P'}}_{{T^*}}}$ is used as the initial spectrum for denoising to obtain the enhanced spectrum ${{\bf{P'}}_0}$. However, in this process, we face a trade-off between the enhancement effectiveness and the consistency between ${{\bf{P'}}_0}$ and ${{\bf{P}}_{zr}}$. If the selected ${T^*}$ is too large, ${{\bf{P'}}_0}$ may deviate from ${{\bf{P}}_{zr}}$. On the other hand, if ${T^*}$ is too small, the noise ${\bm{\kappa }}$ may remain in ${{\bf{P'}}_0}$, which could affect the subsequent sensing tasks. To handle the issues mentioned above, we use ${{\bf{P'}}_c}$ as a condition to guide the generation process, thereby encouraging the purified spectrum to approximate ${{\bf{P'}}_c}$ while reducing the noise. Specifically, using ${{\bf{P'}}_c}$ as the generation guidance, each transition is sampled according to 
    \begin{align}\label{eq25}
{p_{\theta ',\gamma }}\left( {{{{\bf{P'}}}_{t - 1}}|{{{\bf{P'}}}_t},{{{\bf{P'}}}_c}} \right) = Z{p_{\theta '}}\left( {{{{\bf{P'}}}_{t - 1}}|{{{\bf{P'}}}_t}} \right){p_\gamma }\left( {{{{\bf{P'}}}_c}|{{{\bf{P'}}}_{t - 1}}} \right),
    \end{align}
where $Z$ is a normalizing constant~\cite{dhariwal2021diffusion}. In practice, it is challenging to sample precisely from (\ref{eq25}), but it can be approximated as a perturbed Gaussian distribution. Hence, based on the principles of the diffusion model described previously, we have
    \begin{align}\label{eq26}
{p_{\theta '}}\left( {{{{\bf{P'}}}_{t - 1}}|{{{\bf{P'}}}_t}} \right) = {\cal N}\left( {{{\bf{P}}_{t - 1}};{{\bf{\mu }}_{\theta '}},{{\bf{\Sigma }}_{\theta '}}} \right).
    \end{align}
Applying the logarithm on both sides of (\ref{eq26}), we can further obtain
    \begin{align}\label{eq27}
\log \left[ {{p_{\theta '}}\left( {{{{\bf{P'}}}_{t - 1}}|{{{\bf{P'}}}_t}} \right)} \right] =  - \frac{1}{2}{\left( {{\bf{P'}} - {{\bf{\mu }}_{\theta '}}} \right)^{\rm{T}}}{\bf{\Sigma }}_{\theta '}^{ - 1}\left( {{\bf{P'}} - {{\bf{\mu }}_{\theta '}}} \right) + C,
    \end{align}
where $C$ is a constant. Assuming that $\log \left[ {{p_\gamma }\left( {{{{\bf{P'}}}_c}|{{{\bf{P'}}}_{t - 1}}} \right)} \right]$ has lower curvature compared with ${\bf{\Sigma }}_{\theta '}^{ - 1}$, the Taylor series expansion around ${{\bf{P'}}_{t - 1}}$ is used to approximate $\log \left[ {{p_\gamma }\left( {{{{\bf{P'}}}_c}|{{{\bf{P'}}}_{t - 1}}} \right)} \right]$, which yields
    \begin{align}\label{eq28}
\log \left[ {{p_\gamma }\left( {{{{\bf{P'}}}_c}|{{{\bf{P'}}}_{t - 1}}} \right)} \right] &\buildrel\textstyle.\over= \log \left[ {{p_\gamma }\left( {{{{\bf{P'}}}_c}|{{{\bf{P'}}}_{t - 1}}} \right)} \right]\left| {_{{{{\bf{P'}}}_{t - 1}} = {{\bf{\mu }}_{\theta '}}}} \right.\\ \notag
 &+ \left( {{{{\bf{P'}}}_{t - 1}} - {{\bf{\mu }}_{\theta '}}} \right)\\ \notag
 &\times {\nabla _{{{{\bf{P'}}}_{t - 1}}}}\log \left[ {{p_\gamma }\left( {{{{\bf{P'}}}_c}|{{{\bf{P'}}}_{t - 1}}} \right)} \right]\left| {_{{{{\bf{P'}}}_{t - 1}} = {{\bf{\mu }}_{\theta '}}}} \right.
    \end{align}
Based on (\ref{eq27}) and (\ref{eq28}), we have
        \begin{align}\label{eq29}
&\log \left[ {{p_{\theta ',\gamma }}\left( {{{{\bf{P'}}}_{t - 1}}|{{{\bf{P'}}}_t},{{{\bf{P'}}}_c}} \right)} \right]\\ \notag
 &\buildrel\textstyle.\over=  - \frac{1}{2}{\left( {{\bf{P'}} - {{\bf{\mu }}_{\theta '}}} \right)^{\rm{T}}}{\bf{\Sigma }}_{\theta '}^{ - 1}\left( {{\bf{P'}} - {{\bf{\mu }}_{\theta '}}} \right) + \left( {{\bf{P'}} - {{\bf{\mu }}_{\theta '}}} \right)\psi  + C'\\ \notag
 &=  - \frac{1}{2}{\left( {{\bf{P'}} - {{\bf{\mu }}_{\theta '}} - {{\bf{\Sigma }}_{\theta '}}\psi } \right)^{\rm{T}}}{\bf{\Sigma }}_{\theta '}^{ - 1}\left( {{\bf{P'}} - {{\bf{\mu }}_{\theta '}} - {{\bf{\Sigma }}_{\theta '}}\psi } \right)\\ \notag
 &+ \frac{1}{2}{\psi ^{\rm{T}}}{{\bf{\Sigma }}_{\theta '}}\psi  + C'\\ \notag
&= \log \left[ {p\left( {\bf{z}} \right)} \right] + C'',
    \end{align}
where 
    \begin{align}\label{eq30}
    {\bf{z}} \sim {\cal N}\left( {{\bf{z}};{{\bf{\mu }}_{\theta '}} + {{\bf{\Sigma }}_{\theta '}}\psi ,{{\bf{\Sigma }}_{\theta '}}} \right),
    \end{align}
    \begin{align}\label{eq31}
    \psi  = {\nabla _{{{{\bf{P'}}}_{t - 1}}}}\log \left[ {{p_\gamma }\left( {{{{\bf{P'}}}_c}|{{{\bf{P'}}}_{t - 1}}} \right)} \right],
    \end{align}
and $C''$ is a constant related to $Z$.

From the derivation, one can see that ${p_{\theta '}}\left( {{{{\bf{P'}}}_{t - 1}}|{{{\bf{P'}}}_t}} \right)$ in (\ref{eq26}) represents an unconditional diffusion model trained on the noise-free AJ spectrum, while ${p_\gamma }\left( {{{{\bf{P'}}}_c}|{{{\bf{P'}}}_{t - 1}}} \right)$ denotes the probability that ${{\bf{P'}}_{t - 1}}$ ultimately denoises to approximate the spectrum ${{\bf{P'}}_c}$. This probability can be approximated as
    \begin{align}\label{eq32}
    {p_\gamma }\left( {{{{\bf{P'}}}_c}|{{{\bf{P'}}}_{t - 1}}} \right) = \frac{1}{{Z'}}\exp \left[ { - s'D\left( {{{{\bf{P'}}}_{t - 1}},{{\bf{P'}}_c}} \right)} \right],
    \end{align}
where $Z'$ is the normalizing factor, $s'$ is a scaling factor that controls the strength of guidance, and $D\left(  \cdot  \right)$ represents the distance evaluator. Here, the mean square error (MSE) is used as the distance metric. In (\ref{eq32}), reducing $D\left( {{{{\bf{P'}}}_{t - 1}},{{\bf{P'}}_c}} \right)$ can effectively increases ${p_\gamma }\left( {{{{\bf{P'}}}_c}|{{{\bf{P'}}}_{t - 1}}} \right)$, thereby encouraging the generated AJ spectrum to approach ${{\bf{P'}}_c}$. By taking the logarithm of both sides of (\ref{eq32}), we can further obtain
\begin{align}\label{eq33}
\log \left[ {{p_\gamma }\left( {{{{\bf{P'}}}_c}|{{{\bf{P'}}}_{t - 1}}} \right)} \right] =  - \log \left( {Z'} \right) - s'D\left( {{{{\bf{P'}}}_{t - 1}},{{\bf{P'}}_c}} \right).
    \end{align}
On this basis, the gradient can be calculated as  
    \begin{align}\label{eq34}
{\nabla _{{{{\bf{P'}}}_{t - 1}}}}\log \left[ {{p_\gamma }\left( {{{{\bf{P'}}}_c}|{{{\bf{P'}}}_{t - 1}}} \right)} \right] =  - s'{\nabla _{{{{\bf{P'}}}_{t - 1}}}}D\left( {{{{\bf{P'}}}_{t - 1}},{{\bf{P'}}_c}} \right).
    \end{align}
From (\ref{eq32}) to (\ref{eq34}), we observe that the conditional transition can be approximated by a Gaussian similar to the unconditional transition operator, but with its mean shifted by $ - s'{{\bf{\Sigma }}_{\theta '}}{\nabla _{{{{\bf{P'}}}_{t - 1}}}}D\left( {{{{\bf{P'}}}_{t - 1}},{{\bf{P'}}_c}} \right)$, thereby enabling the generated AJ spectrum to approach ${{\bf{P'}}_c}$ with less noise.

 In the aforementioned process, $s'$ is the key factor influencing guidance. If $s'$ is large, it can effectively guide the diffusion model to generate a noise-free AJ spectrum. However, from another perspective, a large $s'$ may lead the generated AJ spectrum to be overly close to the condition ${{\bf{P'}}_c}$, thereby preserving the noise from ${{\bf{P'}}_c}$. Consequently, we set $s'$ to a value related to ${t^*}$.  When ${t^*} - 1$ is large, the noise in ${{\bf{P'}}_{{t^*} - 1}}$ has already been disrupted during the diffusion process, and hence a larger $s'$ can be used to guide the generation of the purified AJ spectrum without the concern of retaining noise from ${{\bf{P'}}_c}$. In cases where ${t^*} - 1$ is small, ${{\bf{P'}}_c}$ still retains some noise, necessitating a smaller $s'$ to control the strength of guidance, thereby preventing the final generated AJ spectrum from retaining noise. Hence, according to (\ref{eq24}), we have
\begin{align}\label{eq35}
{{{\bf{P'}}}_{{t^*}}} &= \sqrt {{{\bar \alpha '}_{{t^*}}}} \left( {{{\bf{P}}_{zr}} + {\bm{\kappa }}} \right) + \sqrt {1 - {{\bar \alpha '}_{{T^*}}}} {{\bm{\varepsilon }}_0}\\ \notag
 &= \sqrt {{{\bar \alpha '}_{{t^*}}}} {{\bf{P}}_{zr}} + \sqrt {{{\bar \alpha '}_{{t^*}}}} {\bm{\kappa }} + \sqrt {1 - {{\bar \alpha '}_{{T^*}}}} {{\bm{\varepsilon }}_0}.
\end{align}
Based on (\ref{eq24}) and (\ref{eq35}), the noise is constrained within the range of $\left[ { - \sqrt {{{\bar \alpha '}_{{t^*}}}} \gamma ,\sqrt {{{\bar \alpha '}_{{t^*}}}} \gamma } \right]$, as the ${l_\infty }$ norm of ${\bm{\kappa }}$ can be bounded to $\gamma $. Meanwhile, $\sqrt {1 - {{\bar \alpha '}_{{t^*}}}} {{\bf{\varepsilon }}_0}$ can approximately be bounded within $\left[ { - 3\sqrt {1 - {{\bar \alpha '}_{{t^*}}}} ,3\sqrt {1 - {{\bar \alpha '}_{{t^*}}}} } \right]$, given that the 99.7\% confidence interval of the standard Gaussian distribution is [-3,3]~\cite{raftery1995hypothesis}. Based on the above analysis, we have 
\begin{align}\label{eq36}
s' = \frac{{3\sqrt {1 - {{\bar \alpha '}_{{t^*}}}} }}{{\gamma \sqrt {{{\bar \alpha '}_{{t^*}}}} }} \cdot a,
\end{align}
where $a$ is the empirically selected parameter based on the image resolution and distance metric. We summarize the AJ spectrum enhancement process in Algorithm 1. It is important to note that, to ensure the effectiveness, for each spectrum, Algorithm 1 performs $W$ rounds of enhancementd.
    \begin{figure*}[htp]
    \centering
    \includegraphics[width=0.9\textwidth]{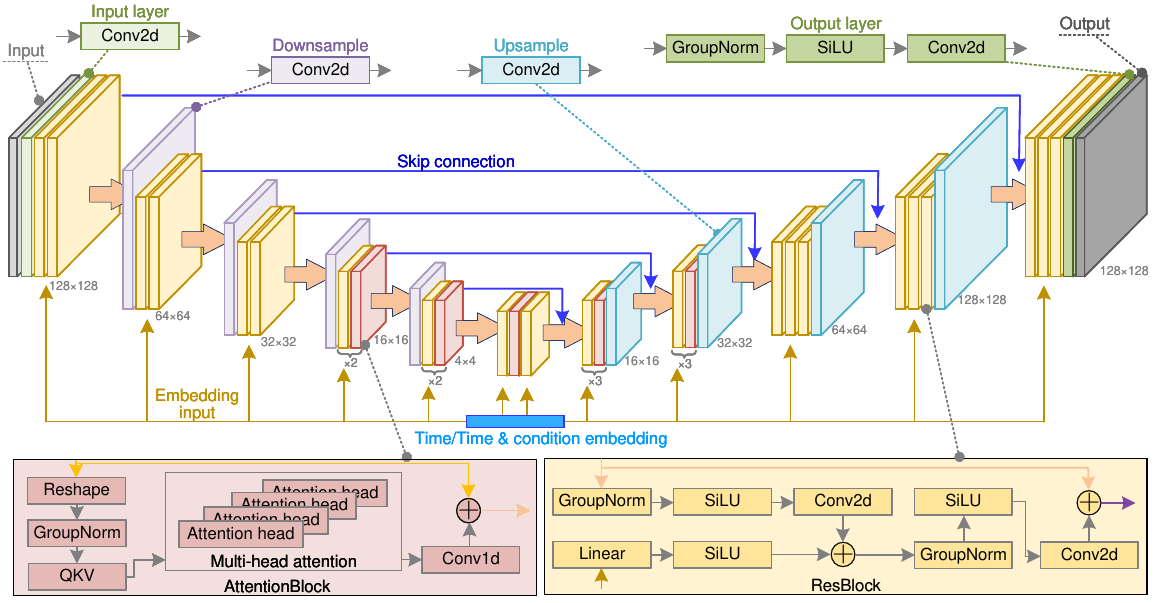}
    \caption{The network architecture of the diffusion model used in DiRA. Note that in the process of sample quantity expansion, DiRA employs conditional generation, hence the embedding includes both time and conditions. In the process of sample quality enhancement, however, the embedding includes only time. SiLU indicates the sigmoid gated linear unit. QKV is the abbreviation of query, key, and value.}
    \label{PURE}
    \end{figure*}
\begin{algorithm}[t]
{\small \caption{Spectrum Quality Enhancement}}
\hspace*{0.02in} {\bf{{Input: Distance gradient, gradient scale $s'$, ${{{\bf{P'}}}_c}$}}}
\begin{algorithmic}[1]
\For {$i \leftarrow 1$ to $W$}
\State Forward process ${{\bf{P'}}_{{t^*}}} = \sqrt {{{\bar \alpha '}_{{t^*}}}} \left( {{{{\bf{P'}}}_c}} \right) + \sqrt {1 - {{\bar \alpha '}_{{t^*}}}} {{\bm{\varepsilon }}_0}$
\For {${t^*} \leftarrow {T^*}$ to 1}
\State $\mu ',{\bf{\Sigma '}} \leftarrow {{\bf{\mu }}_{\theta '}}\left( {{{{\bf{P'}}}_{{t^*}}}} \right){{\bf{\Sigma }}_{\theta '}}\left( {{{{\bf{P'}}}_{{t^*}}}} \right)$
\State Sample ${\cal N}\left( {{\bf{\mu '}} - s'{{\bf{\Sigma }}_{\theta '}}{\nabla _{{{{\bf{P'}}}_{t - 1}}}}D\left( {{{{\bf{P'}}}_{t - 1}},{{{\bf{P'}}}_{{t^*} - 1}}} \right),{\bf{\Sigma }}} \right) $ to obtain ${{\bf{P'}}_{{t^*} - 1}}$ 
\EndFor
\State \textbf{end for}
\EndFor
\State \textbf{end for}
\State \Return The enhanced spectrum ${{\bf{P'}}_0}$
\label{AG1}
\end{algorithmic}
\end{algorithm}

\subsection{Target Detection}
Given that the number of peak points in the AJ spectrum is equal to the number of dynamic targets, we train a ResNet based classifier using the augmented AJ spectra to detect the number of targets, thereby validating the effectiveness of DiRA in ISAC networks.

Specifically, we first expand the quantity of AJ spectra. In practice, due to the uneven distribution of collected AJ spectra across different categories, we employ a conditional diffusion model to equalize the sample sizes across these categories. For example, in the case of detecting the number of dynamic targets, suppose that 1,000 original spectra can be collected when there are three targets, while only 100 spectra are collected when there are ten targets. Then, DiRA expands the spectra of both categories to 2,000. It is important to note that the sample size expansion is conducted for all categories, which ensures that each category has both generated and collected spectra. Building on this, DiRA further enhances the quality of the spectra. Similar to quantity expansion, to maintain consistency, both collected and generated spectra need to be enhanced. Finally, using the enhanced spectra, we train a ResNet-based classifier to recognize AJ spectra, thereby detecting the number of dynamic targets.

 ResNet is a deep network constructed by using residual units, with the core being the residual blocks. In this paper, we employ the ResNet-18 model, which consists of 1 convolutional layer, 2 pooling layers, 8 residual units, and 1 fully connected layer. Each residual unit includes two 3x3 convolutional layers and a fully connected layer, with the ReLU function used as the activation function in the convolutional layers. The diagram in Fig.~\ref{RES} illustrates the structure of the model. After training, the model can classify newly collected AJ spectra, thereby improving the detection accuracy.

    \begin{figure}[t]
    \centering
    \includegraphics[width=0.4\textwidth]{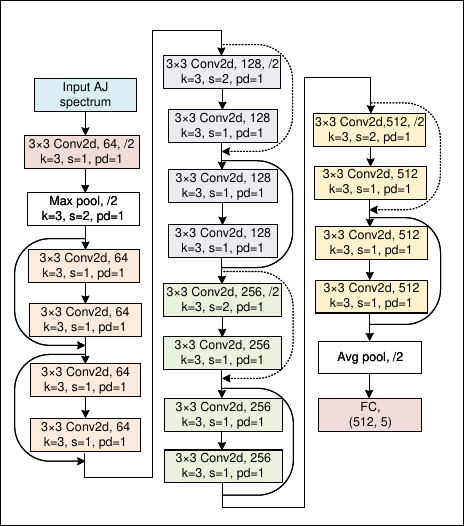} 
    \caption{The structure of the ResNet-18 model. In the model, $k$ is the kernel size, $s$ indicates the stride, and $pd$ represents the padding. } 
    \label{RES} 
    
    \end{figure}

    \begin{figure*}[htp]
    \centering
    \includegraphics[width=1\textwidth]{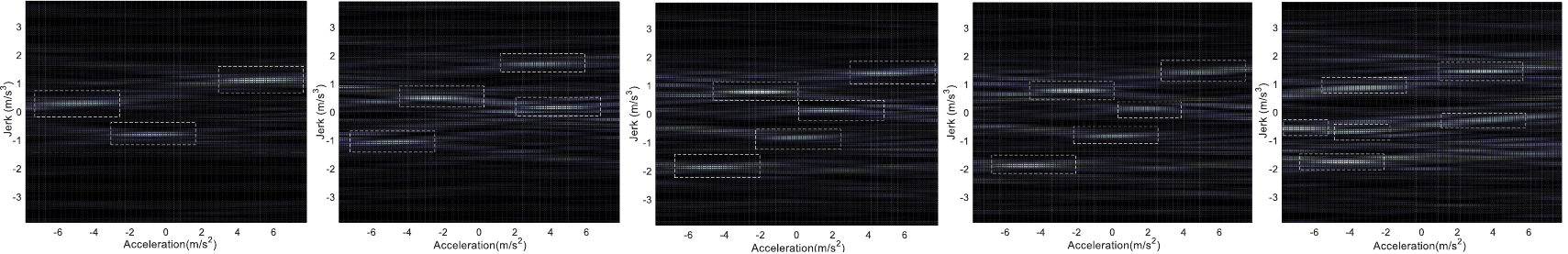}
    \caption{The AJ spectra extracted from the CSI collected from scenarios with different number of targets. }
    \label{SPC}
    \vspace{-0.3cm}
    \end{figure*}
     
    \section{IMPLEMENTATION AND EVALUATION}
    This section provides a thorough evaluation of the proposed DiRA, covering three key aspects. First, we validate the proposed acceleration and jerk estimation algorithms to ensure that the calculated AJ spectrum can effectively supports the sensing function in ISAC networks. Next, we analyze the performance of DiRA in enhancing both the quantity and quality of the  AJ spectra. Finally, we train a ResNet using the augmented dataset and conduct experiments on detecting the number of targets with the trained ResNet, so as to analyze the effectiveness and robustness of the augmented data in supporting ISAC networks.
    
    \subsection{Experimental Configurations}
    We use multiple commercial access points based on the IEEE 802.11 protocol to collect CSI measurement~\cite{gringoli2019free}. The platform features four radio frequency channels, supports up to 4x4 MIMO transmission, and covers both 2 GHz and 5 GHz frequency bands. To minimize noise, directional antennas are employed to help signal transmission and reception. During the experiments, the signal frequency is set to 5.805 GHz, with a bandwidth of 80 MHz, covering 256 subcarriers. The collected data is processed on a server, which included extracting the AJ spectrum, training diffusion models, and generating the new AJ spectrum. The server is equipped with Ubuntu 22.04 operating system, powered by an Intel(R) Xeon(R) Silver 4410Y 12-core processor, and an NVIDIA RTX A6000 GPU.

    \subsection{Experimental Method}
    Regarding the first aspect, we collect CSI in different real-world scenarios with varying numbers of dynamic targets and calculate the AJ spectrum to analyze its effectiveness in detecting the number of targets. For the sample augmentation, we first train a conditional diffusion model on an imbalanced dataset with 150,000 steps and use the trained model to generate 200 new AJ spectra to assess DiRA's performance in sample generation. Next, we set the training steps to 100,000 and train the conditional diffusion model using datasets with different total numbers of samples. After that, another 200 AJ spectra are generated to analyze how the number of original training samples impacts the model's generation performance. During this process, each category within the dataset contains an equal number of samples. Finally, we process the generated AJ spectra to validate DiRA's performance in sample quality enhancement. Here, the number of training steps is 150,000, $T^*$ is set to 50, $W$ is set to 2, and the structural similarity (SSIM) is used as the error distance calculator $D\left(  \cdot  \right)$. Finally, we train the ResNet model using both the original data with quality enhancement and the augmented data. Based on the trained model, we perform the detection of the number of targets in various real-world scenarios, validating the effectiveness and robustness of DiRA through cross-validation. For the above mentioned evaluations, the Frechet Inception Distance (FID) is used as the metric to measure the distribution similarity between the generated and original sample. A lower FID indicates that the generated AJ spectrum is closer to the collected spectrum, demonstrating better data augmentation performance. For target detection, the detection accuracy is employed to evaluate the performance.
    
    \subsection{Results Analysis}
    \subsubsection{Spectrum Analysis}
    First, we collect the CSI in scenarios with 3 to 7 targets and compute the AJ spectrum using the proposed algorithm. As the results in Fig.~\ref{SPC} illustrated, the proposed algorithm can effectively compute the AJ spectrum based on the collected CSI, thereby supporting the acceleration and jerk estimation for different dynamic targets. As each target has different location, movement direction, and speed, the acceleration and jerk differ for each target, which allows us to identify various dynamic targets in the AJ spectrum. Meanwhile, as more targets appear, the AJ spectrum shows more noise, as indicated in the results. This happens because more targets cause more interference with each other and introduce more cross-terms, making the AJ spectrum contains more noise. However, overall, one can still identify the number of targets and the relevant parameters for each target from the AJ spectrum, which provides solid support for ISAC applications such as target detection and tracking.

    \subsubsection{Quantity Enhancement}
\begin{figure*}[htbp]
	\centering
	\subfigure[The AJ spectrum generation process for the scenario with three targets, and its comparison with CGAN.]
    { \begin{minipage}{16cm}
    \includegraphics[width=1\textwidth]{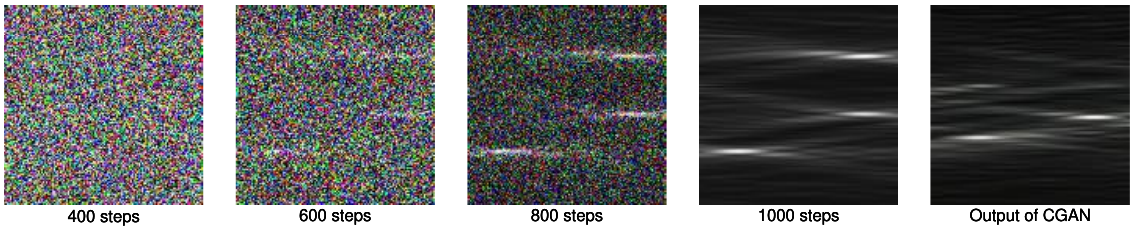}
		\end{minipage}
	}
    \\
	\subfigure[The AJ spectrum generation process for the scenario with seven targets, and its comparison with CGAN. ] {
		\begin{minipage}{16cm}
		\includegraphics[width=1\textwidth]{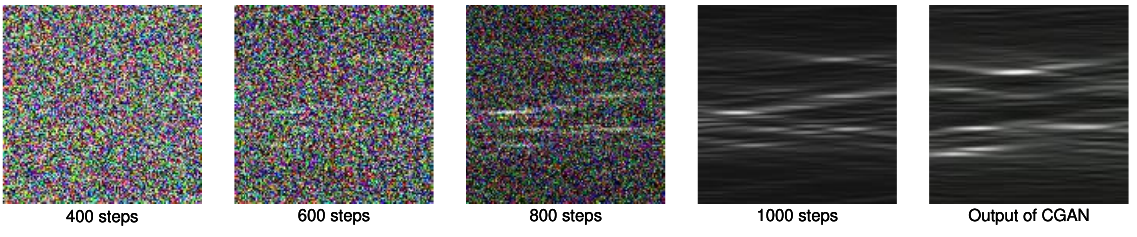}
		\end{minipage}
	}
    \caption{The AJ spectrum generation process, results, and comparison with CGAN across different scenarios.  The x-axis and y-axis represent acceleration and jerk, respectively.}
    \label{GeNe}
    \end{figure*}

    Building on the AJ spectrum analysis, we train a conditional diffusion model and CGAN~\cite{lucic2018gans}, which is widely used in many works~\cite{ye2020deep,patel2020data,tang2023wireless}, to generate new AJ spectra. Then, we compare the spectra generated by these two models to evaluate the generative capabilities of DiRA.  Here, for the five scenarios involving 3 to 7 dynamic targets, the number of training samples are 1800, 1600, 1400, 1200, and 1000 respectively, forming a dataset with  a limited number of samples and uneven distribution. Using the trained model, the generation process is presented in Fig.~\ref{GeNe}. We can see that, even with limited and unevenly distributed training samples, DiRA can effectively generate new AJ spectra through multiple denoising steps based on the input condition, i.e., the number of dynamic targets. These generated spectra, in terms of overall trends and the shapes of spectral peaks, are similar to the original training samples, confirming DiRA's effectiveness and robustness in AJ spectrum generation. 

    Additionally, we observe that in cases with more targets, the generated spectra contain more noise, which can be attributed to two factors. First, as previously analyzed, the AJ spectra inherently have more noise when more targets appear. Second, in this experiment, there are fewer training samples available when there are more targets, which may result in insufficient model training. Comparing the spectra generated by DiRA and CGAN, we can observe that those from DiRA are better, particularly in terms of background noise. This is because the diffusion model can simulate more complex and nonlinear distributions and completes the spectrum generation through multiple iterative denoising steps, resulting in overall better quality. As a result, the peak points corresponding to targets in the AJ spectrum are distinct, making them easier to identify. 

\begin{figure}[t]
    \centering
    \includegraphics[width=0.4\textwidth]{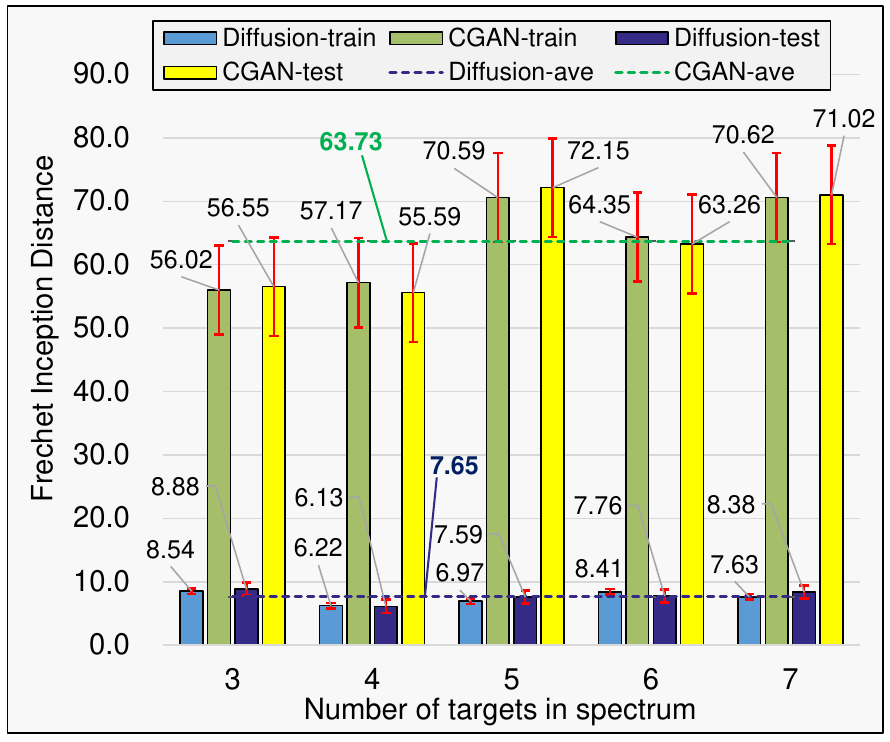} 
    \caption{Comparison between DiRA and CGAN in terms of FID of the AJ spectrum generation.} 
    \label{FIDC} 
    \end{figure}

    After that, we calculate the FID between the generated AJ spectra from DiRA and both the original training and testing spectra, and then compare these results with those from CGAN. As shown in Fig.~\ref{FIDC}, the FID scores of samples generated by DiRA are lower than those of CGAN. Specifically, the average FIDs for the DiRA-generated samples is 7.65, which are lower than CGAN's 63.73. This indicates that the AJ spectra generated by DiRA are closer to the training and testing sets in terms of the distribution, demonstrating better generative capabilities. Additionally, CGAN's FID scores rise significantly with 5, 6, or 7 targets due to uneven training sample distribution, while DiRA remains stable. Overall, DiRA's FID standard deviations are 0.979 for training samples and 1.040 for testing samples, better than CGAN's 7.024 and 7.780. This further demonstrates DiRA's robustness, making it more suitable for data augmentation.

\begin{figure}
	\centering
	\subfigure[The FID between the generated and the training samples. ]
    { \begin{minipage}{7.3cm}
    \includegraphics[width=\textwidth]{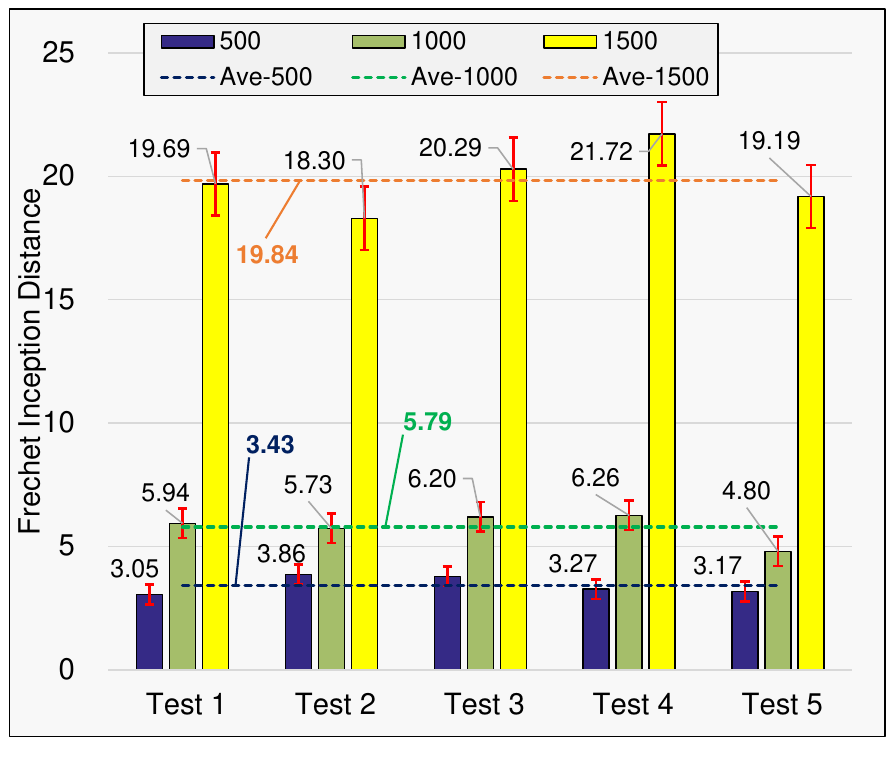}
		\end{minipage}
	}
	\subfigure[The FID between the generated and the testing samples. ] {
		\begin{minipage}{7.3cm}
		\includegraphics[width=\textwidth]{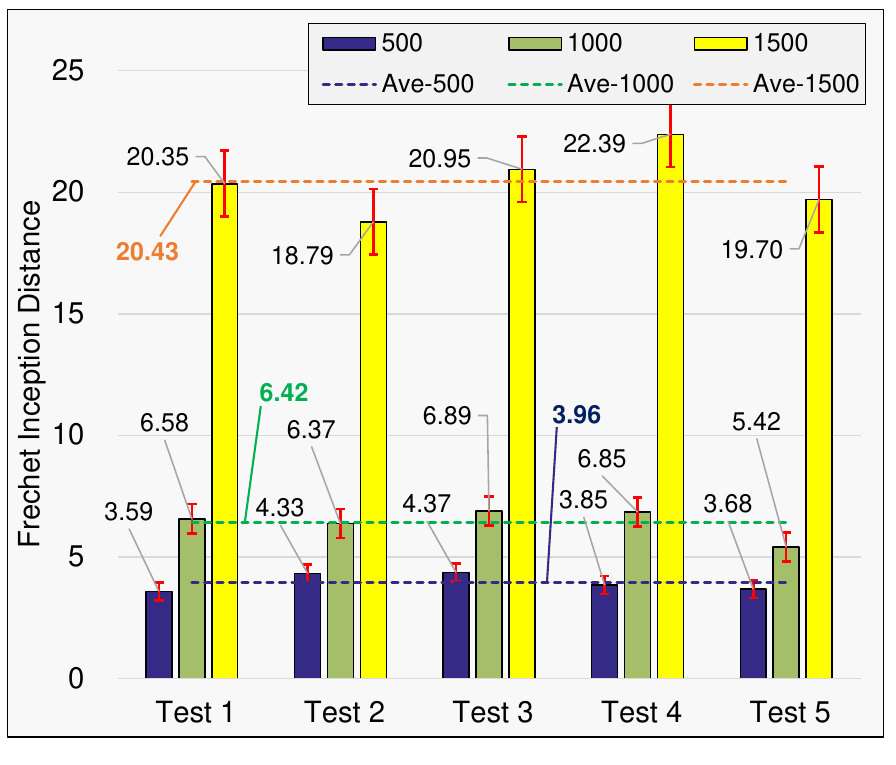}
		\end{minipage}
	}
    \caption{The impact of the original training sample size on DiRA's generative performance.}
    \label{IOR}
    \end{figure}

    Next, we fix the number of training steps at 100,000 and analyze the impact of the original training sample size on DiRA's generative performance. We conduct five independent experiments using the trained DiRA, with the results shown in the Fig.~\ref{IOR}. As shown, for both training and testing samples, the FIDs gradually increase as the size of the training sample grows, given a fixed number of training steps. Specifically, with 500, 1000, and 1500 training samples, the average FIDs of the generated AJ spectra with respect to the training samples are 3.43, 5.79, and 19.84, respectively. A similar trend can be observed for the testing dataset as well. This indicates that with a limited number of training steps, more training data can degrade DiRA's generative performance. This is because, with more training samples, DiRA struggles to fully learn the latent patterns and capture the distribution characteristics within the limited steps. Therefore, in scenarios with limited computing resources, it is necessary to optimize the training sample size appropriately to achieve better generative performance.

    \subsubsection{Quality Enhancement}
    After the quantity expansion, we evaluate DiRA's performance in enhancing the quality of AJ spectra. Figure~\ref{PURE} illustrates the process of DiRA in improving AJ spectrum quality. From the diagram, we can see that DiRA can disrupt the inherent noise in the AJ spectrum by adding noise during the forward process, and then eliminate both the added noise and the inherent noise through denoising. Comparing original and enhanced spectra, it is clear that the enhanced AJ spectrum retain the peaks related to dynamic targets and their locations in the original AJ spectra, demonstrating that DiRA does not lose core information during the enhancement process. Furthermore, the enhanced AJ spectrum exhibits less noise, indicating that DiRA can effectively guide the generation towards a noise-free direction, validating its effectiveness in quality enhancement.

    \begin{figure*}[htp]
    \centering
    \includegraphics[width=0.9\textwidth]{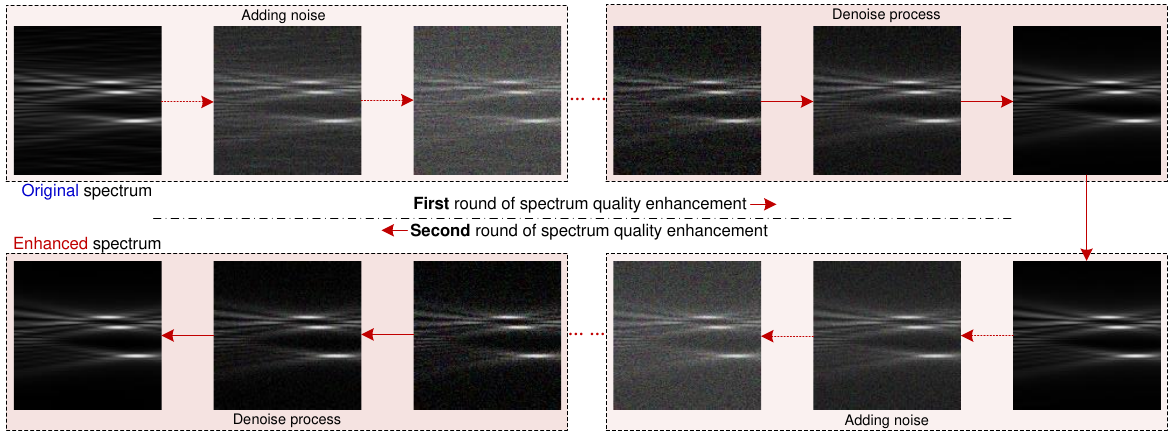}
    \caption{The AJ spectra quality enhancement process. In DiRA, $T^*$ is set to 50, $W$ is set to 2. Therefore, the figure demonstrates two rounds of enhancement processes, each consisting of adding noise and reducing noise. During the noise addition process, the added noise can disrupt the inherent noise within the signal spectra, allowing both the added and inherent noise to be eliminated during the denoising process. It should be noted that the input for the second round of enhancement is the output from the first round.}
    \label{PURE}
    \end{figure*}
    
\begin{figure}[t]
    \centering
    \includegraphics[width=0.4\textwidth]{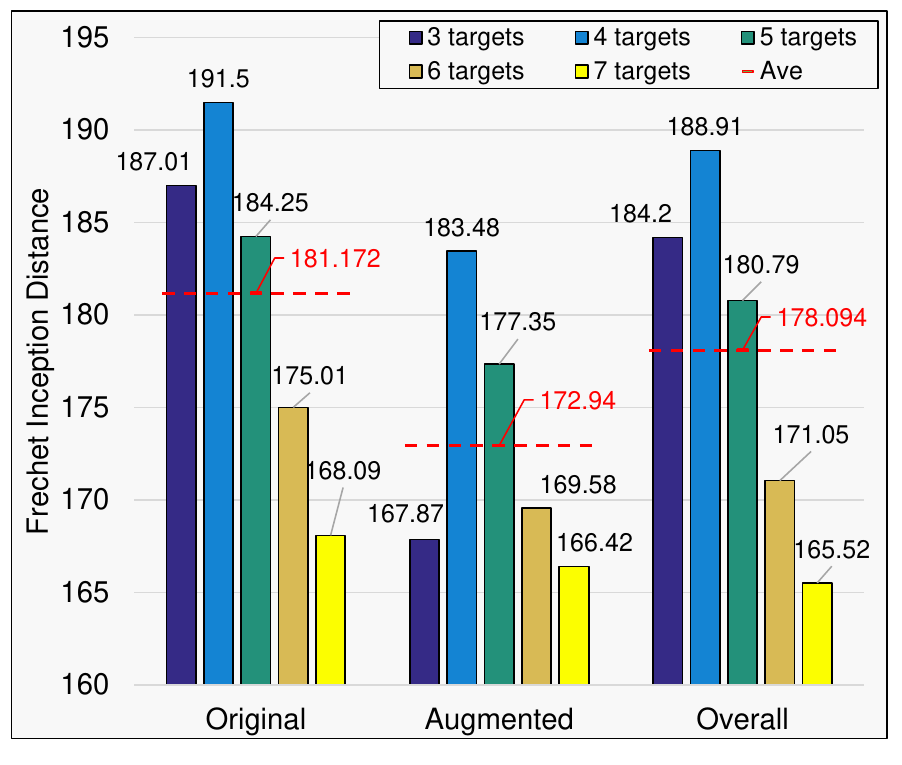} 
    \caption{The FIDs without quality enhancement.} 
    \label{DA-ORG} 
    \end{figure}

\begin{figure}
	\centering
	\subfigure[The FIDs when the spectra are enhanced with $T^* =50 $ and $W=1$.]
    { \begin{minipage}{7.3cm}
    \includegraphics[width=\textwidth]{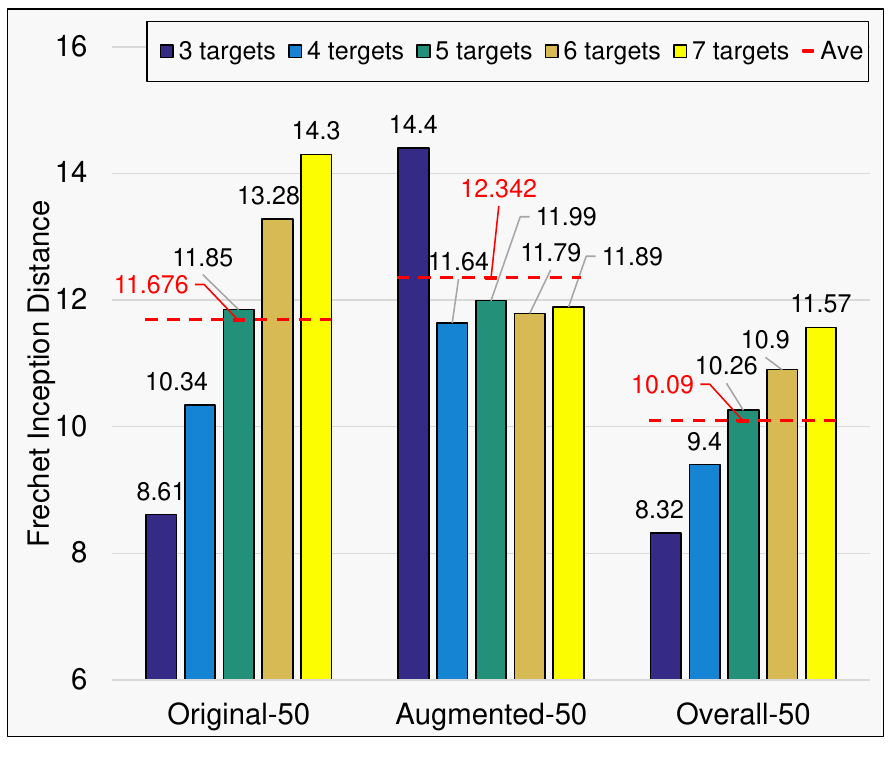}
		\end{minipage}
	}
	\subfigure[The FIDs when the spectra are enhanced with $T^* =100 $ and $W=1$.] {
		\begin{minipage}{7.3cm}
		\includegraphics[width=\textwidth]{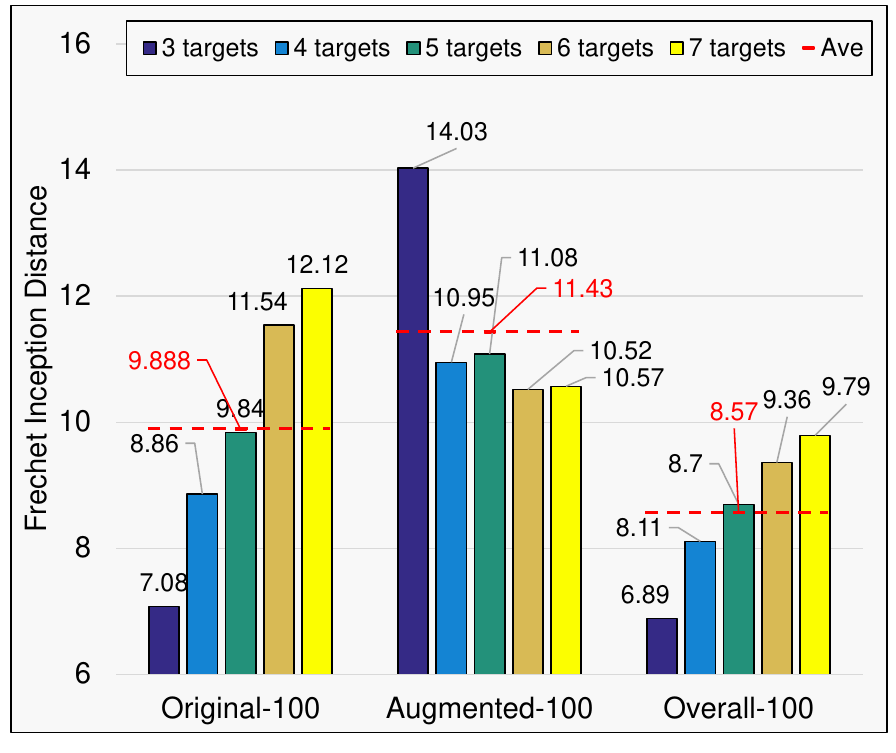} 
		\end{minipage}
	}
	\subfigure[The FIDs when the spectra are enhanced with $T^* =50 $ and $W=2$.] {
		\begin{minipage}{7.3cm}
		\includegraphics[width=\textwidth]{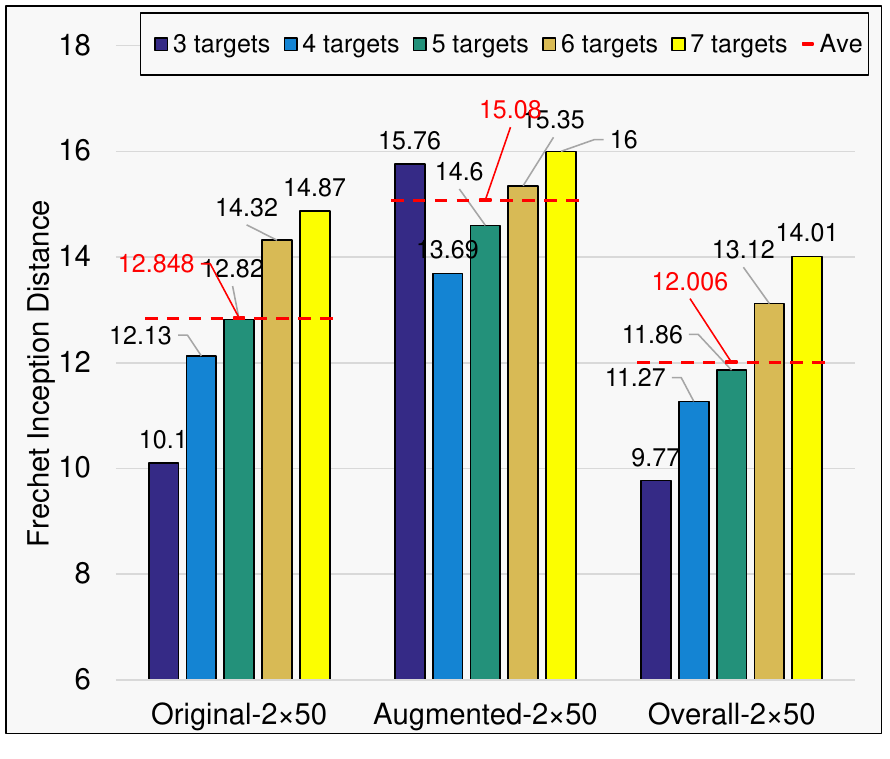} 
		\end{minipage}
	}
 
\caption{The FIDs between the enhanced signal spectra and noise-free signal spectra across various scenarios with a different numbers of targets.}
\label{PFID}
\end{figure}

    Building on this, we calculate the FID for both the original training samples and the DiRA-generated samples before and after quality enhancement, using the noise-free AJ spectra as a reference. As shown in Figs.~\ref{DA-ORG} and \ref{PFID}, the average FIDs for the original samples, generated samples, and overall samples are 181.172, 172.94, and 178.094, respectively before enhancement. After the quality enhancement with $T^* =50 $ and $W=2$, these metrics decrease to 12.848, 15.08, and 12.006, respectively. This indicates that both the original and generated AJ spectra become closer to the noise-free spectra after quality enhancement, further proving DiRA's effectiveness. Additionally, the FIDs of the enhancement with $T^* =50 $ and $W=2$ are higher than 50-step and 100-step enhancement. This result suggests that the method used in DiRA retains certain features from original samples while reducing noise, laying a foundation for improving sensing accuracy in subsequent steps.

    \subsubsection{Target Detection}
    Finally, we train the ResNet-18 model using different datasets and conduct the detection of the number of targets across various scenarios to evaluate DiRA's support for ISAC network sensing capabilities via cross-validation. As illustrated in Fig.~\ref{RES}, the model trained with original data achieves a detection accuracy of 0.64 in the original scenario and 0.32 and 0.28 in two other scenarios. After expanding the quantity of training samples, the accuracy improves to 0.65, 0.48, and 0.45, demonstrating that increasing the size of training data can effectively boost the detection performance. Building on this, a ResNet-18 model trained on DiRA-augmented data reaches detection accuracies of 0.67, 0.71, and 0.71 across three scenarios, indicating an improvement of up to 70\% compared to the case without DiRA. Additionally, compared to enhancements with 50 and 100 steps, DiRA not only achieves high average detection accuracy but also shows less fluctuation across different scenarios, further proving its robustness in data augmentation.
    \begin{figure}[t]
    \centering
    \includegraphics[width=0.4\textwidth]{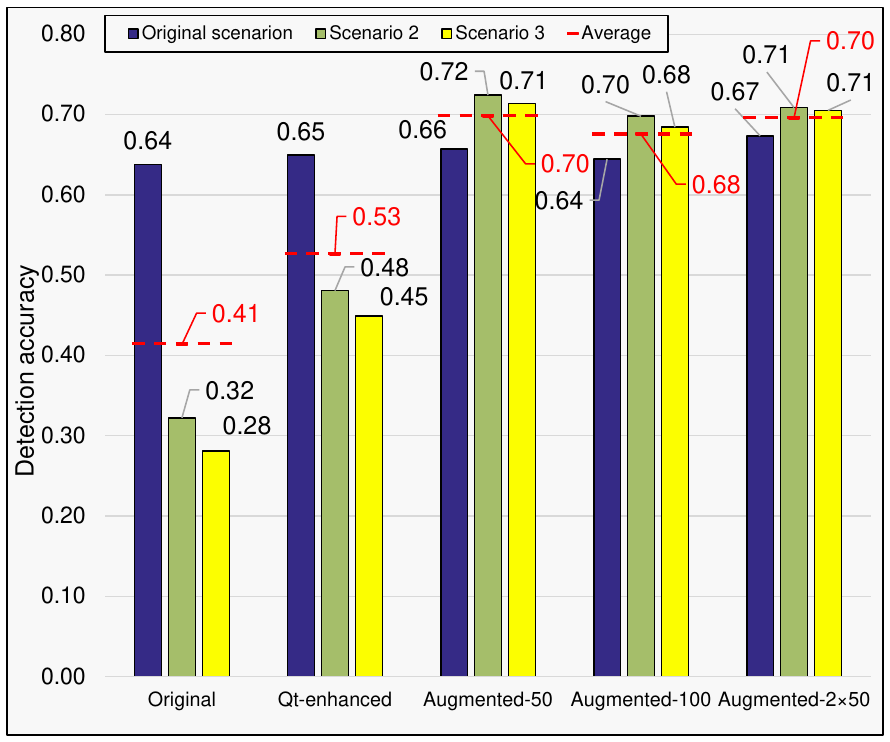} 
    \caption{The cross-validation for detecting the number of targets. } 
    \label{RES} 
    \end{figure}
    \section{Conclusion}
    This paper introduces DiRA, a GenAI-based robust data augmentation framework for ISAC networks. DiRA comprises a sample quantity expansion module and a sample quality enhancement module. We have designed the first module to increase the number of samples in the sensing dataset, while the second module further enhances the quality of these samples, thereby achieving robust data augmentation. Additionally, we have proposed a novel method to extract the acceleration-jerk spectrum from the CSI, which reflects the fluctuation characteristics of multipath signals and can be utilized for various sensing applications such as target detection and behavior recognition. Utilizing DiRA, we have augmented the AJ spectrum and realize the detection of the number of targets based on the augmented spectrum, verifying DiRA's reliability and robustness in enhancing sensing data. This capability provides strong data augmentation support for ISAC networks, effectively addressing the issue of insufficient sensing data. Future work will explore the boundaries of GenAI-based data augmentation in terms of quantity and other aspects.
    
    \bibliographystyle{IEEEtran}
    \bibliography{IEEEabrv,Ref}

\begin{thebibliography}{10}
\providecommand{\url}[1]{#1}
\csname url@samestyle\endcsname
\providecommand{\newblock}{\relax}
\providecommand{\bibinfo}[2]{#2}
\providecommand{\BIBentrySTDinterwordspacing}{\spaceskip=0pt\relax}
\providecommand{\BIBentryALTinterwordstretchfactor}{4}
\providecommand{\BIBentryALTinterwordspacing}{\spaceskip=\fontdimen2\font plus
\BIBentryALTinterwordstretchfactor\fontdimen3\font minus \fontdimen4\font\relax}
\providecommand{\BIBforeignlanguage}[2]{{%
\expandafter\ifx\csname l@#1\endcsname\relax
\typeout{** WARNING: IEEEtran.bst: No hyphenation pattern has been}%
\typeout{** loaded for the language `#1'. Using the pattern for}%
\typeout{** the default language instead.}%
\else
\language=\csname l@#1\endcsname
\fi
#2}}
\providecommand{\BIBdecl}{\relax}
\BIBdecl

\bibitem{liu2022integrated}
F.~Liu, Y.~Cui, C.~Masouros, J.~Xu, T.~X. Han, Y.~C. Eldar, and S.~Buzzi, ``Integrated sensing and communications: Toward dual-functional wireless networks for 6g and beyond,'' \emph{IEEE journal on selected areas in communications}, vol.~40, no.~6, pp. 1728--1767, 2022.

\bibitem{liu2020radar}
F.~Liu, W.~Yuan, C.~Masouros, and J.~Yuan, ``Radar-assisted predictive beamforming for vehicular links: Communication served by sensing,'' \emph{IEEE Transactions on Wireless Communications}, vol.~19, no.~11, pp. 7704--7719, 2020.

\bibitem{liu2020joint}
F.~Liu, C.~Masouros, A.~P. Petropulu, H.~Griffiths, and L.~Hanzo, ``Joint radar and communication design: Applications, state-of-the-art, and the road ahead,'' \emph{IEEE Transactions on Communications}, vol.~68, no.~6, pp. 3834--3862, 2020.

\bibitem{yuan2022orthogonal}
W.~Yuan, Z.~Wei, S.~Li, R.~Schober, and G.~Caire, ``Orthogonal time frequency space modulation—part iii: Isac and potential applications,'' \emph{IEEE Communications Letters}, vol.~27, no.~1, pp. 14--18, 2022.

\bibitem{liu2021cramer}
F.~Liu, Y.-F. Liu, A.~Li, C.~Masouros, and Y.~C. Eldar, ``Cram{\'e}r-rao bound optimization for joint radar-communication beamforming,'' \emph{IEEE Transactions on Signal Processing}, vol.~70, pp. 240--253, 2021.

\bibitem{wang2023through}
J.~Wang, H.~Du, D.~Niyato, M.~Zhou, J.~Kang, Z.~Xiong, and A.~Jamalipour, ``Through the wall detection and localization of autonomous mobile device in indoor scenario,'' \emph{IEEE Journal on Selected Areas in Communications}, 2023.

\bibitem{wang2016csi}
X.~Wang, L.~Gao, S.~Mao, and S.~Pandey, ``Csi-based fingerprinting for indoor localization: A deep learning approach,'' \emph{IEEE transactions on vehicular technology}, vol.~66, no.~1, pp. 763--776, 2016.

\bibitem{wang2024acceleration}
J.~Wang, H.~Du, D.~Niyato, M.~Zhou, J.~Kang, and H.~V. Poor, ``Acceleration estimation of signal propagation path length changes for wireless sensing,'' \emph{IEEE Transactions on Wireless Communications}, 2024.

\bibitem{chen2017tr}
C.~Chen, Y.~Han, Y.~Chen, H.-Q. Lai, F.~Zhang, B.~Wang, and K.~R. Liu, ``Tr-breath: Time-reversal breathing rate estimation and detection,'' \emph{IEEE Transactions on Biomedical Engineering}, vol.~65, no.~3, pp. 489--501, 2017.

\bibitem{nirmal2021deep}
I.~Nirmal, A.~Khamis, M.~Hassan, W.~Hu, and X.~Zhu, ``Deep learning for radio-based human sensing: Recent advances and future directions,'' \emph{IEEE Communications Surveys \& Tutorials}, vol.~23, no.~2, pp. 995--1019, 2021.

\bibitem{liu2019wireless}
J.~Liu, H.~Liu, Y.~Chen, Y.~Wang, and C.~Wang, ``Wireless sensing for human activity: A survey,'' \emph{IEEE Communications Surveys \& Tutorials}, vol.~22, no.~3, pp. 1629--1645, 2019.

\bibitem{zheng2019zero}
Y.~Zheng, Y.~Zhang, K.~Qian, G.~Zhang, Y.~Liu, C.~Wu, and Z.~Yang, ``Zero-effort cross-domain gesture recognition with wi-fi,'' in \emph{Proceedings of the 17th annual international conference on mobile systems, applications, and services}, 2019, pp. 313--325.

\bibitem{he2023robust}
Z.~He, X.~Zhang, Y.~Wang, Y.~Lin, G.~Gui, and H.~Gacanin, ``A robust csi-based wi-fi passive sensing method using attention mechanism deep learning,'' \emph{IEEE Internet of Things Journal}, vol.~10, no.~19, pp. 17\,490--17\,499, 2023.

\bibitem{wang2024aigc}
Z.~Wang, C.~Yang, and S.~Mao, ``Aigc for rf-based human activity sensing,'' \emph{IEEE Internet of Things Journal}, 2024.

\bibitem{wang2024reflexnoop}
P.~Wang, J.~Hu, C.~Liu, and J.~Luo, ``Reflexnoop: Passwords snooping on nlos laptops leveraging screen-induced sound reflection,'' in \emph{Proceedings of the 2024 on ACM SIGSAC Conference on Computer and Communications Security}, 2024, pp. 3361--3375.

\bibitem{wang2025aigc}
Z.~Wang and S.~Mao, ``Aigc for wireless sensing: Diffusion-empowered human activity sensing,'' \emph{IEEE Transactions on Cognitive Communications and Networking}, 2025.

\bibitem{wang2024generativeM}
J.~Wang, H.~Du, D.~Niyato, J.~Kang, S.~Cui, X.~S. Shen, and P.~Zhang, ``Generative ai for integrated sensing and communication: Insights from the physical layer perspective,'' \emph{IEEE Wireless Communications}, 2024.

\bibitem{wang2024unified}
J.~Wang, H.~Du, D.~Niyato, J.~Kang, Z.~Xiong, D.~Rajan, S.~Mao, and X.~Shen, ``A unified framework for guiding generative ai with wireless perception in resource constrained mobile edge networks,'' \emph{IEEE Transactions on Mobile Computing}, 2024.

\bibitem{hussain2021adaptive}
M.~Hussain and N.~Michelusi, ``Adaptive beam alignment in mm-wave networks: A deep variational autoencoder architecture,'' in \emph{2021 IEEE Global Communications Conference (GLOBECOM)}.\hskip 1em plus 0.5em minus 0.4em\relax IEEE, 2021, pp. 1--6.

\bibitem{chi2024rf}
G.~Chi, Z.~Yang, C.~Wu, J.~Xu, Y.~Gao, Y.~Liu, and T.~X. Han, ``Rf-diffusion: Radio signal generation via time-frequency diffusion,'' in \emph{Proceedings of the 30th Annual International Conference on Mobile Computing and Networking}, 2024, pp. 77--92.

\bibitem{dhariwal2021diffusion}
P.~Dhariwal and A.~Nichol, ``Diffusion models beat gans on image synthesis,'' \emph{Advances in neural information processing systems}, vol.~34, pp. 8780--8794, 2021.

\bibitem{trabucco2023effective}
B.~Trabucco, K.~Doherty, M.~Gurinas, and R.~Salakhutdinov, ``Effective data augmentation with diffusion models,'' \emph{arXiv preprint arXiv:2302.07944}, 2023.

\bibitem{croitoru2023diffusion}
F.-A. Croitoru, V.~Hondru, R.~T. Ionescu, and M.~Shah, ``Diffusion models in vision: A survey,'' \emph{IEEE Transactions on Pattern Analysis and Machine Intelligence}, vol.~45, no.~9, pp. 10\,850--10\,869, 2023.

\bibitem{yang2021decimeter}
R.~Yang, X.~Yang, J.~Wang, M.~Zhou, Z.~Tian, and L.~Li, ``Decimeter level indoor localization using wifi channel state information,'' \emph{IEEE Sensors Journal}, vol.~22, no.~6, pp. 4940--4950, 2021.

\bibitem{li2024wifi}
W.~Li, R.~Gao, J.~Xiong, J.~Zhou, L.~Wang, X.~Mao, E.~Yi, and D.~Zhang, ``Wifi-csi difference paradigm: Achieving efficient doppler speed estimation for passive tracking,'' \emph{Proceedings of the ACM on Interactive, Mobile, Wearable and Ubiquitous Technologies}, vol.~8, no.~2, pp. 1--29, 2024.

\bibitem{wang2017phasebeat}
X.~Wang, C.~Yang, and S.~Mao, ``Phasebeat: Exploiting csi phase data for vital sign monitoring with commodity wifi devices,'' in \emph{2017 IEEE 37th international conference on distributed computing systems (ICDCS)}.\hskip 1em plus 0.5em minus 0.4em\relax IEEE, 2017, pp. 1230--1239.

\bibitem{wang2020csi}
------, ``On csi-based vital sign monitoring using commodity wifi,'' \emph{ACM Transactions on Computing for Healthcare}, vol.~1, no.~3, pp. 1--27, 2020.

\bibitem{niu2021understanding}
K.~Niu, F.~Zhang, X.~Wang, Q.~Lv, H.~Luo, and D.~Zhang, ``Understanding wifi signal frequency features for position-independent gesture sensing,'' \emph{IEEE Transactions on Mobile Computing}, vol.~21, no.~11, pp. 4156--4171, 2021.

\bibitem{zhou2021indoor}
M.~Zhou, Y.~Li, H.~Yuan, J.~Wang, and Q.~Pu, ``Indoor wlan personnel intrusion detection using transfer learning-aided generative adversarial network with light-loaded database,'' \emph{Mobile Networks and Applications}, vol.~26, pp. 1024--1042, 2021.

\bibitem{zhang2024vawss}
C.~Zhang, Y.~Zhang, J.~Zhou, and D.~Yuan, ``Vawss: Variational autoencoder-enhanced wireless sensing simulator for wifi channel state information,'' in \emph{Companion of the 2024 on ACM International Joint Conference on Pervasive and Ubiquitous Computing}, 2024, pp. 106--110.

\bibitem{kompella2024augmenting}
S.~K. Kompella, K.~Davaslioglu, Y.~E. Sagduyu, and S.~Kompella, ``Augmenting training data with vector-quantized variational autoencoder for classifying rf signals,'' in \emph{MILCOM 2024-2024 IEEE Military Communications Conference (MILCOM)}.\hskip 1em plus 0.5em minus 0.4em\relax IEEE, 2024, pp. 1--6.

\bibitem{zou2020adversarial}
H.~Zou, C.-L. Chen, M.~Li, J.~Yang, Y.~Zhou, L.~Xie, and C.~J. Spanos, ``Adversarial learning-enabled automatic wifi indoor radio map construction and adaptation with mobile robot,'' \emph{IEEE Internet of Things Journal}, vol.~7, no.~8, pp. 6946--6954, 2020.

\bibitem{zhao2021gsmac}
Y.~Zhao, C.~Liu, K.~Zhu, S.~Zhang, and J.~Wu, ``Gsmac: Gan-based signal map construction with active crowdsourcing,'' \emph{IEEE Transactions on Mobile Computing}, vol.~22, no.~4, pp. 2190--2204, 2021.

\bibitem{wang2024generative}
J.~Wang, H.~Du, D.~Niyato, Z.~Xiong, J.~Kang, B.~Ai, Z.~Han, and D.~I. Kim, ``Generative artificial intelligence assisted wireless sensing: Human flow detection in practical communication environments,'' \emph{IEEE Journal on Selected Areas in Communications}, 2024.

\bibitem{djurovic2012hybrid}
I.~Djurovic, M.~Simeunovic, S.~Djukanovic, and P.~Wang, ``A hybrid cpf-haf estimation of polynomial-phase signals: Detailed statistical analysis,'' \emph{IEEE Transactions on Signal Processing}, vol.~60, no.~10, pp. 5010--5023, 2012.

\bibitem{lv2011lv}
X.~Lv, G.~Bi, C.~Wan, and M.~Xing, ``Lv's distribution: principle, implementation, properties, and performance,'' \emph{IEEE Transactions on Signal Processing}, vol.~59, no.~8, pp. 3576--3591, 2011.

\bibitem{lv2009keystone}
X.~Lv, M.~Xing, S.~Zhang, and Z.~Bao, ``Keystone transformation of the wigner--ville distribution for analysis of multicomponent lfm signals,'' \emph{Signal Processing}, vol.~89, no.~5, pp. 791--806, 2009.

\bibitem{raftery1995hypothesis}
A.~E. Raftery, W.~Gilks, S.~Richardson, and D.~Spiegelhalter, ``Hypothesis testing and model,'' \emph{Markov chain Monte Carlo in practice}, vol.~1, pp. 165--87, 1995.

\bibitem{gringoli2019free}
F.~Gringoli, M.~Schulz, J.~Link, and M.~Hollick, ``Free your csi: A channel state information extraction platform for modern wi-fi chipsets,'' in \emph{Proceedings of the 13th International Workshop on Wireless Network Testbeds, Experimental Evaluation \& Characterization}, 2019, pp. 21--28.

\bibitem{lucic2018gans}
M.~Lucic, K.~Kurach, M.~Michalski, S.~Gelly, and O.~Bousquet, ``Are gans created equal? a large-scale study,'' \emph{Advances in neural information processing systems}, vol.~31, 2018.

\bibitem{ye2020deep}
H.~Ye, L.~Liang, G.~Y. Li, and B.-H. Juang, ``Deep learning-based end-to-end wireless communication systems with conditional gans as unknown channels,'' \emph{IEEE Transactions on Wireless Communications}, vol.~19, no.~5, pp. 3133--3143, 2020.

\bibitem{patel2020data}
M.~Patel, X.~Wang, and S.~Mao, ``Data augmentation with conditional gan for automatic modulation classification,'' in \emph{Proceedings of the 2nd ACM Workshop on wireless security and machine learning}, 2020, pp. 31--36.

\bibitem{tang2023wireless}
H.~Tang, Y.~Zhao, G.~Wang, C.~Luo, and W.~Wang, ``Wireless signal denoising using conditional generative adversarial networks,'' in \emph{IEEE INFOCOM 2023-IEEE Conference on Computer Communications Workshops (INFOCOM WKSHPS)}.\hskip 1em plus 0.5em minus 0.4em\relax IEEE, 2023, pp. 1--6.

\end{thebibliography}


    \end{document}